\documentclass[11pt,a4paper]{article}
\pdfoutput=1 % to allow pdflatex compilation in JCAP
\usepackage{graphicx}
\usepackage{xcolor}
\usepackage[caption=false]{subfig}
\usepackage{mathrsfs,mathtools}
\usepackage{physics,amssymb}
\usepackage{bm}
\usepackage{braket}
\usepackage{listings}
\usepackage{cases}
\usepackage{comment}
\usepackage{soul}
\usepackage{cancel}
\usepackage{cases}
\usepackage[utf8]{inputenc}
\usepackage{url}
\usepackage{empheq}
\usepackage{longtable}
\usepackage[normalem]{ulem}
\usepackage{xspace}
\usepackage{aas_macros}
\usepackage{setspace}
\usepackage{bigints}
\usepackage{jcappub}

\definecolor{DARKBLUE}{HTML}{00008b}
\hypersetup{colorlinks=true
,urlcolor=DARKBLUE
,anchorcolor=DARKBLUE
,citecolor=DARKBLUE
,filecolor=DARKBLUE
,linkcolor=DARKBLUE
,menucolor=DARKBLUE
,linktocpage=true
,pdfproducer=medialab
}

\allowdisplaybreaks

\makeatletter
\gdef\@fpheader{ }
\g@addto@macro\bfseries{\boldmath}
\makeatother

\let\oldsqrt\sqrt
\def\sqrt{\mathpalette\DHLhksqrt}
\def\DHLhksqrt#1#2{%
\setbox0=\hbox{$#1\oldsqrt{#2\,}$}\dimen0=\ht0
\advance\dimen0-0.2\ht0
\setbox2=\hbox{\vrule height\ht0 depth -\dimen0}%
{\box0\lower0.4pt\box2}}

\newcommand{\ee}{e}

\newcommand{\boldmathsymbol}[1]{{\ensuremath{\boldsymbol{#1}}}}

\newcommand{\umin}{\mathrm{min}}

\newcommand{\calP}{\mathcal{P}}

\newcommand{\GeV}{\mathrm{GeV}}

\newcommand{\beq}{\begin{equation}}
\newcommand{\eeq}{\end{equation}}
\newcommand{\bea}{\begin{equation}\begin{aligned}}
\newcommand{\eea}{\end{aligned}\end{equation}}

\newlength{\wsingfig}
\newlength{\wdblefig}
\newlength{\wquadfig}
\newlength{\wtriplefig}
\newcommand{\Eq}[1]{Eq.~(\ref{#1})}
\newcommand{\Eqs}[1]{Eqs.~(\ref{#1})}
\newcommand{\Fig}[1]{Fig.~{\ref{#1}}}

\newcommand{\Refa}[1]{Ref.~{\cite{#1}}}
\newcommand{\Refs}[1]{Refs.~{\cite{#1}}}
\newcommand{\Sec}[1]{Sec.~\ref{#1}}
\newcommand{\Secs}[1]{Secs.~\ref{#1}}
\newcommand{\App}[1]{Appendix~\ref{#1}}
\newcommand{\Apps}[1]{Appendices~\ref{#1}}

\newcommand{\ie}{i.e.\@\xspace}

\title{Real-space Bell inequalities in de Sitter}
\date{\today}

\author[a]{Lloren\c c Espinosa-Portal\'es,}
\affiliation[a]{Instituto de F\'isica Te\'orica UAM-CSIC, Universidad Auton\'oma de Madrid,
Cantoblanco, 28049 Madrid, Spain}
\emailAdd{llorenc.espinosa@uam.es}
\author[b]{Vincent Vennin}
\affiliation[b]{Laboratoire Astroparticule et Cosmologie, CNRS Universit\'e de Paris, \\
10 rue Alice Domon et L\'eonie Duquet, 75013 Paris, France}
\emailAdd{vincent.vennin@cnrs.fr}

\abstract{Bell-inequality violations reveal the presence of quantum correlations between two particles that have interacted and then separated. Their generalisation to quantum fields is necessary to study a number of field-theoretic setups, such as cosmological density fluctuations. In this work, we show how Bell operators can be constructed for quantum fields in real space, and for Gaussian states we compute their expectation value in terms of the field power spectra. We then apply our formalism to a scalar field in de-Sitter space-time. We find that, in spite of the tremendous production of entangled particles with opposite wave momenta on large scales, Bell inequalities are not violated in real space. The reason is that, when considering measurements of a field at two distinct locations in real space, one implicitly traces over the configuration of the field at every other location, leading to a mixed bipartite system. This ``effective decoherence'' effect is responsible for the erasure of quantum features, and casts some doubts on our ability to reveal the quantum origin of cosmological structures. We finally discuss these results in the light of quantum discord.}

\begin{document}

\maketitle
\flushbottom
\section{Introduction}
\label{sec:intro}
Bell-inequality~\cite{Bell:1964kc} violations show that quantum systems display correlations that cannot be accounted for by using a local (\ie no physical influences can travel faster than the speed of light) and realist (\ie physical systems possess complete sets of properties prior to measurement) description. They have thus played a key role in experimentally distinguishing between the quantum predictions and classical properties~\cite{Freedman:1972zza}.

Although Bell inequalities were originally constructed for pairs of spin particles, their generalisation to field-theoretic setups would allow us to study a number of situations where the manifestation of purely quantum effects in fields remains to be investigated. This is notably the case in cosmology, where the gravitational amplification of quantum vacuum fluctuations~\cite{Starobinsky:1979ty,Mukhanov:1981xt, Hawking:1982cz, Starobinsky:1982ee, Guth:1982ec, Bardeen:1983qw} during an era of early accelerated expansion called inflation~\cite{Starobinsky:1980te, Sato:1980yn, Guth:1980zm, Linde:1981mu, Albrecht:1982wi, Linde:1983gd} gives rise to all structures observed in the universe. Although this mechanism leads to predictions that are in exquisite agreement with observations~\cite{Akrami:2018odb}, it takes place at energy scales that can be as high as $10^{16}\, \GeV$, where quantum mechanics has never been tested so far; and it relies on quantising fluctuations of the metric, which may not be conceptually trivial. This is why an experimental test (such as Bell-inequality violation) for the quantum origin of cosmological structures would bring valuable insight~\cite{Campo:2005sv, Maldacena:2015bha, Choudhury:2016cso, Martin:2016tbd, Martin:2017zxs, Kanno:2017dci}.

The goal of this paper is thus to construct Bell operators for quantum fields. In practice, our proposal is then applied to scalar fields in a de-Sitter background, since it will shed direct light on the cosmological context mentioned above. However, the generic framework we outline is applicable to other field-theoretic contexts. 

Let us start by recalling some of the previous results that have been obtained in the direction that we further explore here. If a quantum field $\phi(\vec{x},t)$ is placed in a homogeneous and isotropic space-time, the translational invariance of the background implies that Fourier modes $\phi_{\vec{k}}$ decouple at linear order (\ie working with quadratic Hamiltonians), hence the wavefunctional of the field can be factorised into
\bea
\label{eq:Psi:Fourier}
\Psi\left[\phi(\vec{x},t)\right]=\prod_{\vec{k}\in\mathbb{R}^{3+}} \Psi_k \left(\phi_{\vec{k}}\right) .
\eea
Here, we have assumed that the field is real, which implies that $\phi_{-\vec{k}}=\phi_{\vec{k}}^*$ and this explains why the Hilbert space is labelled by half of the real set only, $\mathbb{R}^{3+}=\mathbb{R}^2\times \mathbb{R}^+$. Since $\phi_{\vec{k}}$ is complex, each Fourier subspace contains two real degrees of freedom, hence it constitutes a bipartite system. A first question concerns the nature of the correlations between sub-parts of this system. 

This can be characterised by means of quantum discord~\cite{Henderson:2001,Zurek:2001}, which is a measure of non-classical correlations that can be applied to cosmological fluctuations~\cite{Lim:2014uea, Martin:2015qta, Hollowood:2017bil}. A computation of quantum discord within a given Fourier subspace of a quantum field was carried out in \Refa{Martin:2015qta}, and although the result depends on the precise way the system is partitioned (see \Refa{Martin:2021znx} for a comprehensive discussion of partition dependence), it was found that quantum discord is not vanishing as soon as the state departs from the vacuum state. This can be understood as follows. Because of space-time isotropy, whenever a particle with momentum $\vec{k}$ is created, it must be entangled with a particle of momentum $-\vec{k}$. Therefore, each Fourier subspace contains entangled particles, and thus possesses a non-vanishing entanglement entropy. Now, because the state is factorisable in Fourier space, see \Eq{eq:Psi:Fourier}, each Fourier subspace is placed in a pure state (described by the wavefunction $\Psi_{\vec{k}}$). Given that quantum discord reduces to half the entanglement entropy for pure states, this explains why a non-vanishing quantum discord is obtained. In practice, for the field describing cosmological perturbations, it was thus found that quantum discord becomes large on super-Hubble scales, suggesting the possible presence of genuinely quantum effects.

For this reason, Bell inequalities within a given Fourier subspace were then studied in \Refs{Martin:2016tbd, Martin:2017zxs}, where it was found that violations can indeed be obtained. When constructing Bell operators in this context, two technical obstacles needed to be overcome. First, Bell operators are combinations of spin operators, hence they are applicable to spin particles. Instead, we are presently dealing with the amplitude of a scalar-field Fourier mode, which is a continuous variable. One thus has to construct pseudo-spin operators (\ie operators satisfying the SU(2) algebra) for continuous variables. Second, if the Hamiltonian is quadratic then the quantum state of the field is Gaussian, which implies that its Wigner function is positive. As shown in \Refa{2005PhRvA..71b2103R} (see also \Refa{Martin:2019wta} for a recent discussion of this result in historical context), the violation of Bell inequalities with non-negative Wigner functions require the use of improper operators, \ie operators whose Wigner-Weyl transform take values outside their spectrum. Such improper pseudo-spin operators have been proposed in \Refs{PhysRevLett.82.2009, PhysRevLett.88.040406, 2004PhLA..324..415G, 2004PhRvA..70b2102L, 2005PhRvA..71b2103R}, and they will be used here too. \\

One of the crucial assumptions behind Bell inequalities is the one of locality, which only makes sense in real space. As a consequence, the violation of Bell inequalities in Fourier space does not necessarily point towards the presence of quantum correlations -- it could be simply due to the fact that physics is not local in Fourier space. Therefore, although the above results are encouraging, they must be translated into real space.

\begin{figure}
    \centering
    \includegraphics[width=0.6\textwidth]{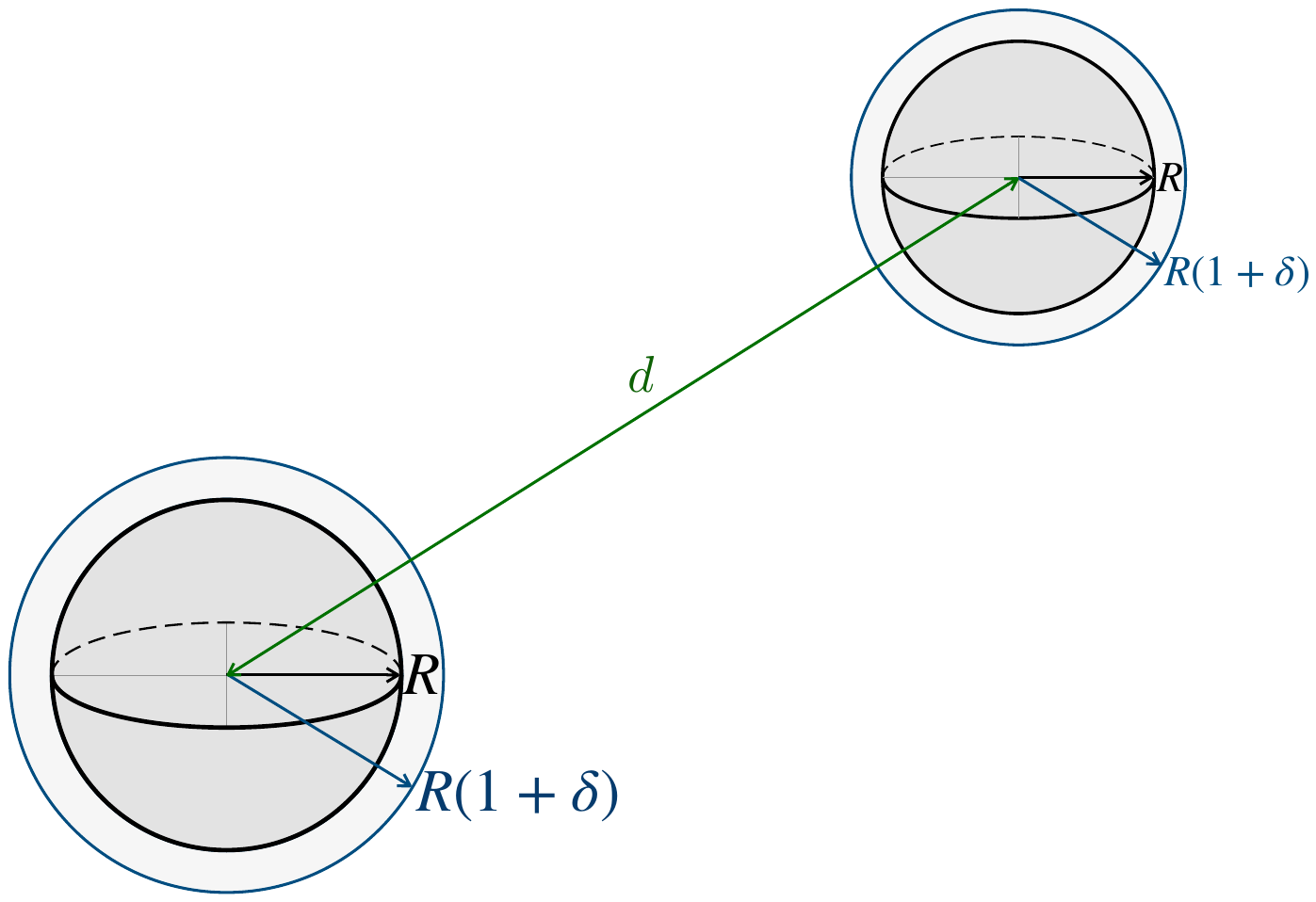}
    \caption{Sketch of the setup studied in this work. A quantum field is coarse grained within two non-overlapping spheres of radius $R$ and distant by $d$. Pseudo-spin measurements are performed in each sphere and correlations between the results are arranged into a combination subject to Bell inequalities.
    %Figure adapted from \Refa{Martin:2021xml}.
    }
    \label{fig:sketch}
\end{figure}
This was done for entanglement entropy and quantum discord in \Refs{Espinosa-Portales:2019peb, Espinosa-Portales:2020pjp, Martin:2021xml, Martin:2021znx}, where the correlations between the field configuration in two non-overlapping spheres in real space were studied. It was found that, although quantum discord is still non vanishing, the typical values encountered are much smaller than those obtained in Fourier space. This can be understood as follows. If the quantum state is factorisable in Fourier space, see \Eq{eq:Psi:Fourier}, then it is not factorisable in real space, which essentially means that the field is subject to non-vanishing spatial correlations. As a consequence, when considering the bipartite system made of the configuration of the field within two spheres, one implicitly traces over the field configuration everywhere outside those two spheres, to which the bipartite system is entangled. Therefore, the quantum state of the bipartite system is a mixed one. This implies that there is no simple relationship between entanglement entropy and quantum discord anymore~\cite{2010PhRvL.105b0503G}. Moreover, this reduction to a mixed state, which was dubbed ``effective decoherence'', is associated with a partial erasure of some of the quantum correlations, which here translates into the reduction of quantum discord. 

This analysis suggests that quantum correlations may be more deeply concealed in real space than in Fourier space. However, since real-space quantum discord is not strictly vanishing, the question of whether real-space correlations display quantum features or not is still open, and the goal of this work is to investigate this issue by constructing real-space Bell inequalities for quantum fields.

Note that there are other obstacles one needs to overcome when trying to reveal the quantum origin of primordial structures~\cite{Martin:2017zxs}. For instance, it is not always clear how one could measure pseudo-spin operators in practice, especially since some of them involve the momentum operator, which in cosmology is associated to the so-called ``decaying mode''. There is also the issue of environmental decoherence (not to be confused with the mechanism of ``effective decoherence'' mentioned above), by which the phase coherence leaks into unobserved degrees of freedom. Let us also mention that, in usual Bell experiments, repeated measurements of the spin correlators have to be performed. In cosmology this can only be done by observing the sky at different locations, and the cosmic variance arising from this effect needs to be incorporated~\cite{Morse:2020mdc}. In this work, we leave these issues aside, and focus on the limitations arising from the behaviour of quantum correlations in real space. Would these limitations be addressed, those other issues would be next in line.

The rest of the paper is organised as follows. In \Sec{sec:Bell:formalism}, we present our generic framework for  real-space Bell inequalities, and apply it to the case of a scalar quantum field placed in a Gaussian state. This leads to an expression for the expectation value of the Bell operator solely in terms of the power spectra of the field. In \Sec{sec:Minkowski}, we apply this formalism to the case of a scalar field in the vacuum state of flat space time, and in \Sec{sec:DeSitter} we carry out a similar analysis in a de-Sitter background. In \Sec{sec:Other:Spin:Operators}, we generalise our findings to another class of pseudo-spin operators. In \Sec{sec:Conclusion} we present our main conclusions, and mention a few possible directions to go beyond the type of Bell tests we presently consider. The paper ends with two technical appendices, \Apps{sec:GKMR:Spin:Correlators} and~\ref{sec:Larsson:Spin:Correlators}, to which some of the technical details of the calculation are deferred, and \App{app:Additional:Figures}, in which we provide additional figures that are not essential to the main discussion but complete our parameter-space exploration.

\section{Bell inequalities for quantum fields}
\label{sec:Bell:formalism}

In this section, we explain how Bell operators can be constructed for quantum fields. Usually, Bell inequalities are considered in the context of dichotomic measurements performed on bipartite systems. Two observers perform measurements of a set of variables $\mathcal{S}_a$ (for instance, spin variables, where ``$a$'' labels the direction of the polariser) for two subsystems 1 and 2 at separate spatial locations $\vec{x}_1$ and $\vec{x}_2$. Upon repeated measurements, they reconstruct the correlators $E(a,b)=\langle \hat{\mathcal{S}}_a(\vec{x}_1) \hat{\mathcal{S}}_b(\vec{x}_2)\rangle$. Under the assumptions of realism and locality, the inequality 
\bea
B=E(a,b)+ E(a,b') + E(a',b) - E(a', b') \leq 2
\eea 
holds, which is known as the Bell inequality~\cite{Bell:1964kc, Clauser:1969ny}. A violation of the Bell inequality thus provides evidence for the presence of genuine quantum effects.

Quantum fields are a priori far from those systems. First, they do not describe pairs of space-like separated objects (they rather ``fill'' the entire space). Second, the associated phase space is described by continuous rather than dichotomic variables. A two-step procedure must therefore be followed, in which two subsystems 1 and 2 are first defined in terms of field measurements at two distinct spatial locations $\vec{x}_1$ and $\vec{x}_2$, and spin-like observables are then constructed out of those continuous-variable measurements.
\subsection{Bipartite systems for two-point measurements of a quantum field}
\label{sec:Bipartite}
Here we summarise the proposal made in \Refa{Martin:2021qkg} to cast two-point measurements of a quantum field in terms of a quantum bipartite system. Only the main arguments and results are given, while further details can be found in \Refa{Martin:2021qkg}. Let $\phi(\vec{x})$ be a real scalar quantum field (since all measurements are performed at the same time, the time argument is omitted for notation convenience) and $\pi(\vec{x})$ its conjugated momentum. We define the field coarse-grained at a location $\vec{x}$ over a radius $R$ as
\bea
\label{eq:def:CoarseGrain}
\phi_{R} (\vec{x}) \equiv \left(\frac{a}{R}\right)^3
\int\dd^3\vec{y} \, \phi (\vec{y})\, 
W\left(\frac{a\left\vert \vec{y} - \vec{x}\right\vert}{R} \right),
\eea
with a similar expression for $\pi_{R}(\vec{x})$. In this formula, $a$ is the scale factor of the universe, such that space is labelled by comoving coordinates,\footnote{This is to make the formalism directly applicable to cosmology in \Sec{sec:DeSitter}. Otherwise, in flat space-time, one may simply set $a=1$ and use physical coordinates, as in \Sec{sec:Minkowski}.} and $W$ is a window function that asymptotes a constant at small arguments and decays at large arguments. It is normalised such that $ \int_0^\infty z^2 W(z)\dd z = 1/(4\pi)$, \ie such that a uniform field is left invariant by the coarse-graining procedure. Moreover, in order for the field and its momentum to commute when coarse-grained around distant spatial locations, the support of $W$ must be taken as compact. In practice, we assume that $W(z)=0$ for $z\geq 1+\delta$, \ie $\delta$ is a parameter that controls the size of the support of $W$ [see \Eq{eq:WindowFunction:Improved} below].
Let us then consider two spatial points $\vec{x}_1$ and $\vec{x}_2$ distant by $d\equiv a \vert \vec{x}_1 - \vec{x}_2 \vert>2R(1+\delta)$ (the situation is summarised in \Fig{fig:sketch}). One has $[\phi_R(\vec{x}_1),\phi_R(\vec{x}_2)]=[\pi_R(\vec{x}_1),\pi_R(\vec{x}_2)]=0$, while thanks to the compactness of $W$, the canonical commutation relation $[\phi(\vec{x}) , \pi(\vec{y})] = i \delta(\vec{x}-\vec{y})$ ensures that
\bea
 \label{eq:commutator:Interm}
 \left[ \phi_R(\vec{x}_i), \pi_R(\vec{x}_j)\right]
 = i 4\pi \left(\frac{a}{R}\right)^3 \int_0^{1+\delta} \dd z  W^2\left(z\right)
 \delta_{ij} \equiv i \frac{3}{4\pi} \left(\frac{a}{R}\right)^3 G\, \delta_{ij}\, ,
\eea 
which defines the parameter $G$, and where the prefactor is arranged such that $G=1$ for a constant window function with $\delta=0$. As a consequence, canonical commutation relations for the coarse-grained fields are recovered only after rescaling the fields according to 
\bea
\tilde{\phi}_R(\vec{x}) =  \left(\frac{R}{a}\right)
  \sqrt{\frac{4\pi}{3 G(\delta)}}{\phi}_R(\vec{x})
  \qquad\text{and}\qquad 
  \tilde{\pi}_R(\vec{x}) =  \left(\frac{R}{a}\right)^2
  \sqrt{\frac{4\pi}{3 G(\delta)}}{\phi}_R(\vec{x}) .
\eea 
The coarse-grained rescaled fields thus describe a bipartite system with canonical commutation relations, which was the goal of this subsection.
\subsection{Pseudo-spin operators}
Our next task is to introduce pseudo-spin operators for the continuous variables describing those coarse-grained rescaled fields. The construction of spin operators out of continuous variables can be addressed in several ways, see \Refa{Martin:2017zxs} for various proposals. Here, for expliciteness, we consider the so-called Gour-Khanna-Mann-Revzen (GKMR) pseudo-spin operators~\cite{2004PhLA..324..415G, 2005PhRvA..71b2103R}, although another set of spin operators will be discussed in \Sec{sec:Other:Spin:Operators}. 

GKMR operators are built from the eigenstates $\ket{\tilde{\phi}_R(\vec{x})}$ of the coarse-grained field configuration. Let us first introduce the auxiliary states
\bea 
\label{eq:E:O:def}
    \ket{\mathcal{E}(\vec{x})} & = \frac{1}{2} \left[ \ket{\tilde{\phi}_R(\vec{x})} + \ket{-\tilde{\phi}_R(\vec{x})} \right]\\
    \ket{\mathcal{O}(\vec{x})} & = \frac{1}{2} \left[ \ket{\tilde{\phi}_R(\vec{x})} - \ket{-\tilde{\phi}_R(\vec{x})} \right] ,
\eea 
in terms of which the GKMR operators are defined as
\bea
\label{eq:GKMR:def}
    \hat{\mathcal{S}}_x(\vec{x}) & = \int_0^{\infty} \dd\tilde{\phi}_R(\vec{x}) \left[\ket{\mathcal{E}(\vec{x})}\bra{\mathcal{O}(\vec{x})} + \ket{\mathcal{O}(\vec{x})}\bra{\mathcal{E}(\vec{x})}\right]\\
    \hat{\mathcal{S}}_y(\vec{x}) & = i \int_0^{\infty} \dd\tilde{\phi}_R(\vec{x}) \left[\ket{\mathcal{O}(\vec{x})}\bra{\mathcal{E}(\vec{x})} - \ket{\mathcal{E}(\vec{x})}\bra{\mathcal{O}(\vec{x})}\right]\\
    \hat{\mathcal{S}}_z(\vec{x}) & = - \int_0^{\infty} \dd\tilde{\phi}_R(\vec{x}) \left[\ket{\mathcal{E}(\vec{x})}\bra{\mathcal{E}(\vec{x})} - \ket{\mathcal{O}(\vec{x})}\bra{\mathcal{O}(\vec{x})}\right]\,.
\eea 
One can check that those operators are indeed pseudo-spin operators, \ie they are such that $\hat{\mathcal{S}}_x^2=\hat{\mathcal{S}}_y^2=\hat{\mathcal{S}}_z^2=1$ and they satisfy the SU(2) commutation relations $[\hat{\mathcal{S}}_x,\hat{\mathcal{S}}_y]=2i\hat{\mathcal{S}}_z$, $[\hat{\mathcal{S}}_y,\hat{\mathcal{S}}_z]=2i\hat{\mathcal{S}}_x$ and $[\hat{\mathcal{S}}_z,\hat{\mathcal{S}}_x]=2i\hat{\mathcal{S}}_y$. 
Note also that since $\hat{\mathcal{S}}_z$ is diagonal in the field eigenvector basis, it involves the field operator only. On the contrary, $\hat{\mathcal{S}}_x$ and $\hat{\mathcal{S}}_y$ are not diagonal, hence they also rely on the momentum operator. As mentioned in \Sec{sec:intro}, in a cosmological context, it implies that their measurement requires access to the decaying mode, which is technically challenging.
\subsection{Gaussian states}
In order to perform explicit calculations, let us assume that the quantum field is placed in a Gaussian state. This case is particularly relevant for cosmology, since at linear order in cosmological perturbation theory, cosmological perturbations are indeed placed in such a state. Gaussian states are easy to represent in phase space, since their Wigner function is simply given by a Gaussian function
\bea
\label{eq:Wigner:Gauss}
W_{\hat{\rho}}(\boldmathsymbol{q}) = \frac{\ee^{-\frac{1}{2}\boldmathsymbol{q}^{\mathrm{T}} \boldmathsymbol{\gamma}^{-1}\boldmathsymbol{q}}}{(2\pi)^2\sqrt{\det \boldmathsymbol{\gamma}}}  ,
\eea 
where we have arranged the bipartite phase-space variables into the vector $\boldmathsymbol{q} = (\tilde{\phi}_R(\vec{x}_1), \tilde{\pi}_R(\vec{x}_1), \allowbreak \tilde{\phi}_R(\vec{x}_2), \tilde{\pi}_R(\vec{x}_2))^\mathrm{T}$. Let us recall that the Wigner function is constructed from the density matrix of the state via a Wigner-Weyl transform~\cite{1927ZPhy...46....1W, 1946Phy....12..405G, 1949PCPS...45...99M} 
\bea 
\label{eq:Wigner-Weyl}
W_{\hat{\rho}}(\boldmathsymbol{q}) &=
 \int_{\mathbb{R}^2} \frac{\dd z_1}{2\pi} \frac{\dd z_2}{2\pi}  e^{- i \tilde{\pi}_R(\vec{x}_1) z_1 - i \tilde{\pi}_R(\vec{x}_2) z_2  } \\
&~~\left< \tilde{\phi}_R(\vec{x}_1) + \frac{z_1}{2}, \tilde{\phi}_R(\vec{x}_2) + \frac{z_2}{2} \right|\hat{\rho} \left| \tilde{\phi}_R(\vec{x}_1)- \frac{z_1}{2}, \tilde{\phi}_R(\vec{x}_2) - \frac{z_2}{2}\right> .
\eea 
It is a phase-space function that gives a full representation of the quantum state (given that the density matrix can be recovered from the Wigner function by an inverse Wigner-Weyl transform). 
This representation is particularly convenient for Gaussian states given its simple form~\eqref{eq:Wigner:Gauss}, and we choose to work with it since it makes the following calculations technically much simpler.

In \Eq{eq:Wigner:Gauss}, $\boldmathsymbol{\gamma}$ is the covariance matrix of the system, \ie it is such that~\cite{Martin:2021xml}\footnote{Note that there is a factor $2$ difference in the definition of the covariance matrix with respect to \Refa{Martin:2021xml}.}
\bea
\label{eq:covariance:matrix}
\gamma_{ab} \equiv &  \left\langle \left\lbrace \hat{q}_a, \hat{q}_b\right\rbrace \right\rangle\\
= &\frac{4\pi }{3G}\left(\frac{R}{a}\right)^3
\int \dd\ln k \, \widetilde{W}^2\left(\frac{R}{a}k\right)
 \\ & \times
\left(
\begin{array}{cccc}
   \displaystyle \frac{a}{R} \calP_{\phi\phi}(k) & \calP_{\phi\pi}(k) &
  \displaystyle  \ \frac{a}{R} \calP_{\phi\phi}(k)\,
  {\mathrm{sinc}\left(\frac{kd}{a} \right)}\ &
  \displaystyle \calP_{\phi\pi}(k)\, \mathrm{sinc}
  \left(\frac{k d}{a}\right)\\ \\
  - & \displaystyle  \frac{R}{a} \calP_{\pi\pi}(k)  & \displaystyle
 \frac{a}{R} \calP_{\phi\phi}(k)\, \mathrm{sinc}\left(\frac{k d}{a}\right)& \displaystyle
  \frac{R}{a} \calP_{\pi\pi}(k) \, \mathrm{sinc}\left(\frac{k d}{a}\right)\\ \\
  - & - & \displaystyle\frac{a}{R}  \calP_{\phi\phi}(k)  & \calP_{\phi\pi}(k)
  \\ \\
 -  & - & - &  \displaystyle\frac{R}{a} \calP_{\pi\pi}(k) 
\end{array}
\right)\, ,
\eea 
where $\lbrace \hat{A},\hat{B}\rbrace \equiv (\hat{A}\hat{B}+\hat{B}\hat{A})/2$ denotes the anticommutator. The second expression above casts the result in terms of the Fourier transform of the window function $\widetilde{W}(z)\equiv  4 z^{-3} \allowbreak \int_0^\infty W(u/z)u\sin u\, \dd u$, and in terms of the reduced power spectra of the field and its momentum, defined as $\langle \lbrace \hat{\phi}_{\vec{k}}^\dagger, \hat{\phi}_{\vec{k}'}\rbrace\rangle = 2\pi^2 k^{-3}\mathcal{P}_{\phi\phi}(k) \delta(\vec{k}-\vec{k}')$ with similar expressions for $\mathcal{P}_{\phi\pi}$ and $\mathcal{P}_{\pi\pi}$. Here, the Fourier transform is defined as $\hat{\phi}(\vec{x})=(2\pi)^{-3/2}\int\dd \vec{k} \ee^{-i\vec{k}\cdot\vec{x}}\hat{\phi}(\vec{k})$ with a similar expression for $\hat{\pi}(x)$, and the above definition of the power spectrum assumes that the field is placed in a quantum state that is invariant under spatial translations and rotations, which is the case in cosmology.

Since the covariance matrix is symmetric, only the upper triangular part has been written explicitly. The invariance of the setup under exchanging subsystems 1 and 2 also leads to an additional symmetry of the covariance matrix, such that there are only $6$ independent entries, namely $\gamma_{11}$, $\gamma_{12}$, $\gamma_{22}$, $\gamma_{13}$, $\gamma_{14}$ and $\gamma_{24}$ (if the two coarse-graining radii are different at $\vec{x}_1$ and $\vec{x}_2$ then that symmetry is lost, but the present formalism can still be used, see \Refa{Martin:2021xml}).

It is important to stress that even if the quantum state of the field is pure, the state of the bipartite system we consider is, in general, mixed. This can be seen by computing the purity parameter 
\bea
\label{eq:purity:def}
\mathfrak{p}\equiv \mathrm{Tr}(\hat{\rho}^2)
=\frac{1}{4\sqrt{\det \boldmathsymbol{\gamma}}},
\eea
which equals one for a pure state but is smaller than one otherwise, and where the second expression is valid for Gaussian states. By considering the configuration of the field in a subset of real space only (namely within a distance $R$ of either $\vec{x}_1$ or $\vec{x}_2$), our bipartite system is implicitly constructed by tracing over its configuration at every other location. By doing so, since the field is correlated in real space, one effectively obtains a mixed state, which explains why $\mathfrak{p} < 1$. This ``effective decoherence'' mechanism~\cite{Martin:2021xml, Martin:2021qkg} has the potential to blur the presence of a genuine quantum signal. In \Refs{Martin:2021xml, Martin:2021qkg}, its effect of quantum discord was investigated, while the goal of the present work is to study how it affects Bell-inequality violations.

Let us also note that the tracing-out procedure in the Hilbert space is equivalent to phase-space marginalisation (see Appendix D of \Refa{Colas:2021llj}). As a consequence, the quantum state of the sub-system 1, which describes the field configuration around the location $\vec{x}_1$, can be alternatively obtained by (i) tracing the density matrix of the full bipartite system over the sub-system 2 or (ii) integrating the Wigner function~\eqref{eq:Wigner:Gauss} over $q_3$ and $q_4$. By following this second route, one still obtains a Gaussian Wigner function, the covariance matrix of which is simply given by the upper-left two-by-two sub-block of \Eq{eq:covariance:matrix}. This makes the phase-space representation of Gaussian states particularly convenient.
\subsection{Spin correlators}
\label{sec:SpinCorrelators}
The Wigner-Weyl transform~\eqref{eq:Wigner-Weyl} can be written for any operator $\hat{O}$, which it hereby translates into a phase-space function $W_{\hat{O}}$. In the Wigner representation, its quantum expectation value is given by
\bea
\label{eq:mean:weyl}
\left< \hat{O}\right> = \mathrm{Tr}(\hat{\rho}\,\hat{O}) = (2\pi)^2\int\dd \boldmathsymbol{q} W_{\hat{\rho}}(\boldmathsymbol{q})W_{\hat{O}}(\boldmathsymbol{q}) .
\eea 
The evaluation of the spin correlators $E(a,b)$ thus requires to first compute the Wigner-Weyl transform of the GKMR operators, and then to integrate it against the Wigner function~\eqref{eq:Wigner:Gauss}. In each subspace, one finds~\cite{Martin:2017zxs} 
%[since the sub phase-spaces are two-dimensional, the prefactor $(2\pi)^{-2}$ in \Eq{eq:Wigner-Weyl} needs to be replaced with $(2\pi)^{-1}$]
%
\bea
\label{eq:Weyl:Transform}
	&W_{\hat{\mathcal{S}}_x(\vec{x})}\left[\tilde{\phi}_R(\vec{x}),\tilde{\pi}_R(\vec{x})\right] = \frac{1}{\left(2\pi\right)} \textrm{sign} \left[\tilde\phi_R(\vec{x})\right]\\
	&W_{\hat{\mathcal{S}}_y(\vec{x})}\left[\tilde{\phi}_R(\vec{x}),\tilde{\pi}_R(\vec{x})\right] = - \frac{1}{\left(2\pi\right)} \delta\left[\tilde\phi_R(\vec{x})\right] \mathcal{P}\left[1/\tilde{\pi}_R(\vec{x})\right]\\
	&W_{\hat{\mathcal{S}}_z(\vec{x})}\left[\tilde{\phi}_R(\vec{x}),\tilde{\pi}_R(\vec{x})\right] = - \frac{1}{2}  \delta\left[\tilde{\phi}_R(\vec{x})\right] \delta\left[\tilde{\pi}_R(\vec{x})\right]
\eea 
where $\mathcal{P}$ denotes the principal part. Since $\hat{\mathcal{S}}_i(\vec{x}_1)$ and $\hat{\mathcal{S}}_j(\vec{x}_2)$ act on two separate sectors of the full Hilbert space, where $i$, $j=x$, $y$ or $z$, the Wigner-Weyl transform of their product is simply given by the product of their Wigner-Weyl transforms, \ie
\bea 
W_{\hat{\mathcal{S}}_i(\vec{x}_1)\otimes \hat{\mathcal{S}}_j(\vec{x}_2)}(\boldmathsymbol{q}) = W_{\hat{\mathcal{S}}_i(\vec{x}_1)}(q_1,q_2)W_{\hat{\mathcal{S}}_j(\vec{x}_2)}(q_3,q_4) .
\eea 
From the above expressions, one can thus readily compute the spin correlators in the state~\eqref{eq:Wigner:Gauss}.

In general, one is free to set the directions of the spin measurements in an arbitrary way. It is however common practice to consider the case where $\hat{\mathcal{S}}_a=\hat{\mathcal{S}}_z$, $\hat{\mathcal{S}}_{a'}=\hat{\mathcal{S}}_x$, and $\hat{\mathcal{S}}_b$ and $\hat{\mathcal{S}}_{b'}$ are set in the $(xz)$ plane, \ie $\hat{\mathcal{S}}_b = \sin\theta \hat{\mathcal{S}}_x + \cos\theta \hat{\mathcal{S}}_z$ and $\hat{\mathcal{S}}_{b'} = \sin\theta' \hat{\mathcal{S}}_x + \cos\theta' \hat{\mathcal{S}}_z$. Since $\braket{\hat{\mathcal{S}}_x(\vec{x}_1) \hat{\mathcal{S}}_z(\vec{x}_2)} = 0$ (see \App{sec:GKMR:Spin:Correlators}), upon optimising the polar angles $\theta$ and $\theta'$ such as to get a maximal value for $B$, one obtains
\bea 
\label{eq:bell:Szz:Sxx}
B = 2 \sqrt{\braket{\hat{\mathcal{S}}_z(\vec{x}_1) \hat{\mathcal{S}}_z(\vec{x}_2)}^2+\braket{\hat{\mathcal{S}}_x(\vec{x}_1) \hat{\mathcal{S}}_x(\vec{x}_2)}^2}\, .
\eea 
We are thus left with two spin correlators to compute. The details of this computation, which essentially boils down to performing Gaussian integrals, are deferred to \App{sec:GKMR:Spin:Correlators}. One finds
\bea 
\label{eq:SzSz}
	\braket{\hat{\mathcal{S}}_z(\vec{x}_1) \hat{\mathcal{S}}_z(\vec{x}_2)} & = \frac{1}{4 \sqrt{\det \boldmathsymbol{\gamma}}} = \mathfrak{p}\, ,
\eea 
where one recovers the purity parameter introduced in \Eq{eq:purity:def}. This already indicates that the ``effective decoherence'' mechanism mentioned above leads to a suppression of the expectation value of the Bell operator, since $B=2\sqrt{\mathfrak{p}^2+\braket{\hat{\mathcal{S}}_x(\vec{x}_1) \hat{\mathcal{S}}_x(\vec{x}_2)}^2}$. One also has
\bea 
\label{eq:SxSx}
	\braket{\hat{\mathcal{S}}_x(\vec{x}_1) \hat{\mathcal{S}}_x(\vec{x}_2)} = 
	- \frac{2}{\pi} \arctan\left[ a_{12} (a_{11} a_{22} - a_{12}^2)^{-1/2}\right]\,
\eea 
where $a_{11}$, $a_{12}$ and $a_{22}$ are the entries of the symmetric two-by-two matrix
\bea 
\label{eq:def:a}
\boldmathsymbol{a}
    = (\boldmathsymbol{\gamma}^{-1})^{\phi\phi} - (\boldmathsymbol{\gamma}^{-1})^{\phi \pi} [(\boldmathsymbol{\gamma}^{-1})^{\pi\pi}]^{-1} (\boldmathsymbol{\gamma}^{-1})^{\pi \phi} .
\eea
In this expression, the two overscripts indicate a restriction of the $\boldmathsymbol{\gamma}^{-1}$ matrix to the lines labelled by the first index, and to the columns labelled by the second index (see \App{sec:GKMR:Spin:Correlators} for further details).

Before applying this formalism to concrete examples in \Secs{sec:Minkowski} and~\ref{sec:DeSitter}, a last word is in order regarding the window function. As mentioned above, it needs to be compact for the phase-space operators at two different locations to commute, and the simplest choice is therefore to consider a Heaviside function. However, as pointed out in \Refs{Martin:2021xml, Martin:2021qkg}, such a sharp window function sometimes creates divergences in some of the intermediate quantities being computed, and it is more convenient to consider a continuous window function that is made of a constant piece and of a linear piece:
\bea
\label{eq:WindowFunction:Improved}
W(x)=\frac{3}{4\pi {\cal F}(\delta)}
\begin{cases}
1 \qquad \text{for} \qquad x\le 1\, ,\\
\displaystyle
-\frac{1}{\delta}(x-1)+1 \qquad \text{for}\qquad 1<x\le 1+\delta\, , \\
0 \qquad \text{for}\qquad x>1+\delta \, ,
\end{cases}
\eea
where ${\mathcal{F}}(\delta)=(\delta+2)(\delta ^2+2\delta +2)/4$ is such that the normalisation condition mentioned above is satisfied. With that expression, the Fourier transform of the window function is given by $\widetilde{W}(z)=3
  \{z\sin (z)
  -(1+\delta)z\sin \left[(1+\delta )z\right]
  +2\cos (z)
  -2
  \cos \left[(1+\delta)z\right]\}/[\delta\mathcal{F}(\delta) z^4]$, and one has $  G(\delta) =8 (\delta ^3+5 \delta ^2+10 \delta +10)/  [5 (\delta +2)^2 (\delta ^2+2 \delta +2)^2]$.
\section{Flat space-time}
\label{sec:Minkowski}
Let us first consider the case of a massless scalar field placed in the vacuum state of the Minkowski space-time. This serves two purposes. First, since there are few physical parameters involved to describe this setup, and given that the field mode functions take simple analytical forms, this provides a simple example to discuss the application of the formalism introduced above. Second, the analysis of \Refa{Martin:2021xml} revealed that this situation might be less trivial than it seems. Even though the field remains in its vacuum state and no particle is being created, it was indeed found in \Refa{Martin:2021xml} that quantum discord is not strictly vanishing (unless $\delta=0$). On the one hand, the absence of entangled particles indicates that no Bell-inequality violation should occur in this setup, but on the other hand, the presence of a non-vanishing discord might suggest otherwise and the question thus needs to be settled.
\subsection{Covariance matrix}
After expanding the field into independent Fourier modes, the mode functions in the vacuum state are given by $\phi_{\vec{k}} = \ee^{- i k t}/\sqrt{2k}$ and $\pi_{\vec{k}}=\dot{\phi}_{\vec{k}}=-i \sqrt{k/2}\ee^{-i k t}$, which give rise to the reduced power spectra $\calP_{\phi\phi} = k^2/(4\pi^2)$, $\calP_{\pi\pi}=k^4/(4\pi^2)$ and $\calP_{\phi\pi}=0$. Plugging those expressions into \Eq{eq:covariance:matrix}, the entries of the covariance matrix can be computed and are given by
\bea
\label{eq:gamma:Minkowski}
    \gamma_{11} &= \frac{\mathcal{K}_1(\delta)}{3\pi G(\delta)} \, ,  \qquad
    \gamma_{12} = 0\, , \qquad
    \gamma_{22} = \frac{\mathcal{K}_3(\delta)}{3\pi G(\delta)}\\
    \gamma_{13} &= \frac{ \mathcal{L}_1 (\alpha, \delta)}{3\pi G(\delta)} \, , \qquad
    \gamma_{14} = 0  \, , \qquad
    \gamma_{24} = \frac{\mathcal{L}_3 (\alpha,\delta)}{3\pi G(\delta)}  \, ,
\eea
where we have introduced the integrals
\bea
\label{eq:Integrals:K:L}
    \mathcal{K}_\mu(\delta) & = \int_0^\infty z^\mu \widetilde{W}^2(z) \dd z\\
    \mathcal{L}_\mu(\alpha,\delta) & = \int_0^\infty z^\mu \widetilde{W}^2 \left(z\right) \mathrm{sinc}(\alpha z) \dd z\,,
\eea
which depend on the parameter
\bea
\label{eq:alpha:def}
    \alpha \equiv \frac{d}{R}\, .
\eea 
This parameter measures the distance between the two patches in units of the patch radius, and as explained in \Sec{sec:Bipartite}, it needs to be larger than $2(1+\delta)$ for the two patches not to overlap. By plugging the expression for $\widetilde{W}$ given below \Eq{eq:WindowFunction:Improved} into \Eq{eq:Integrals:K:L}, those integrals can be performed analytically, and the corresponding expressions can be found in \Refa{Martin:2021xml}. We do not reproduce them here since they are not particularly insightful at this stage.
\subsection{Spin and Bell correlators}
Let us now evaluate the GKMR spin correlators following the method outlined in \Sec{sec:SpinCorrelators}. Since the field-momentum correlators vanish, namely $\gamma_{12}=\gamma_{14}=0$, one can show that it is also true for the inverse covariance matrix, \ie $(\boldmathsymbol{\gamma}^{-1})^{\phi\pi}=\boldmathsymbol{0}$. As a consequence, \Eq{eq:def:a} leads to 
\bea 
\boldmathsymbol{a}= (\boldmathsymbol{\gamma}^{-1})^{\phi\phi} = \frac{1}{\gamma_{11}^2-\gamma_{13}^2}
\begin{pmatrix}
\gamma_{11} & -\gamma_{13}\\
-\gamma_{13} & \gamma_{11}
\end{pmatrix}\, ,
\eea
so \Eqs{eq:SxSx} and~\eqref{eq:SzSz} give rise to
\bea  
	\braket{\hat{\mathcal{S}}_x(\vec{x}_1) \hat{\mathcal{S}}_x(\vec{x}_2)} & = \frac{2}{\pi}\mathrm{arctan}\left(\frac{\gamma_{13}}{\sqrt{\gamma_{11}^2-\gamma_{13}^2}}\right)=\frac{2}{\pi}\mathrm{arctan}\left[\frac{\mathcal{L}_1 (\alpha, \delta)}{\sqrt{\mathcal{K}_1^2 (\delta)-\mathcal{L}_1^2 (\alpha, \delta)}}\right]\\
	\braket{\hat{\mathcal{S}}_z(\vec{x}_1) \hat{\mathcal{S}}_z(\vec{x}_2)} & =\frac{1}{4\sqrt{ \left(\gamma_{11}^2-\gamma_{13}^2\right)\left(\gamma_{22}^2-\gamma_{24}^2\right)}}=\frac{9\pi^2G^2(\delta)}{4\sqrt{\left[\mathcal{K}_1^2 (\delta)-\mathcal{L}_1^2 (\alpha, \delta)\right]\left[\mathcal{K}_3^2 (\delta)-\mathcal{L}_3^2 (\alpha, \delta)\right]}}
\eea
where the result is also given in terms of the integrals introduced in \Eq{eq:Integrals:K:L}.
\begin{figure}
    \centering
    \includegraphics[width=\textwidth]{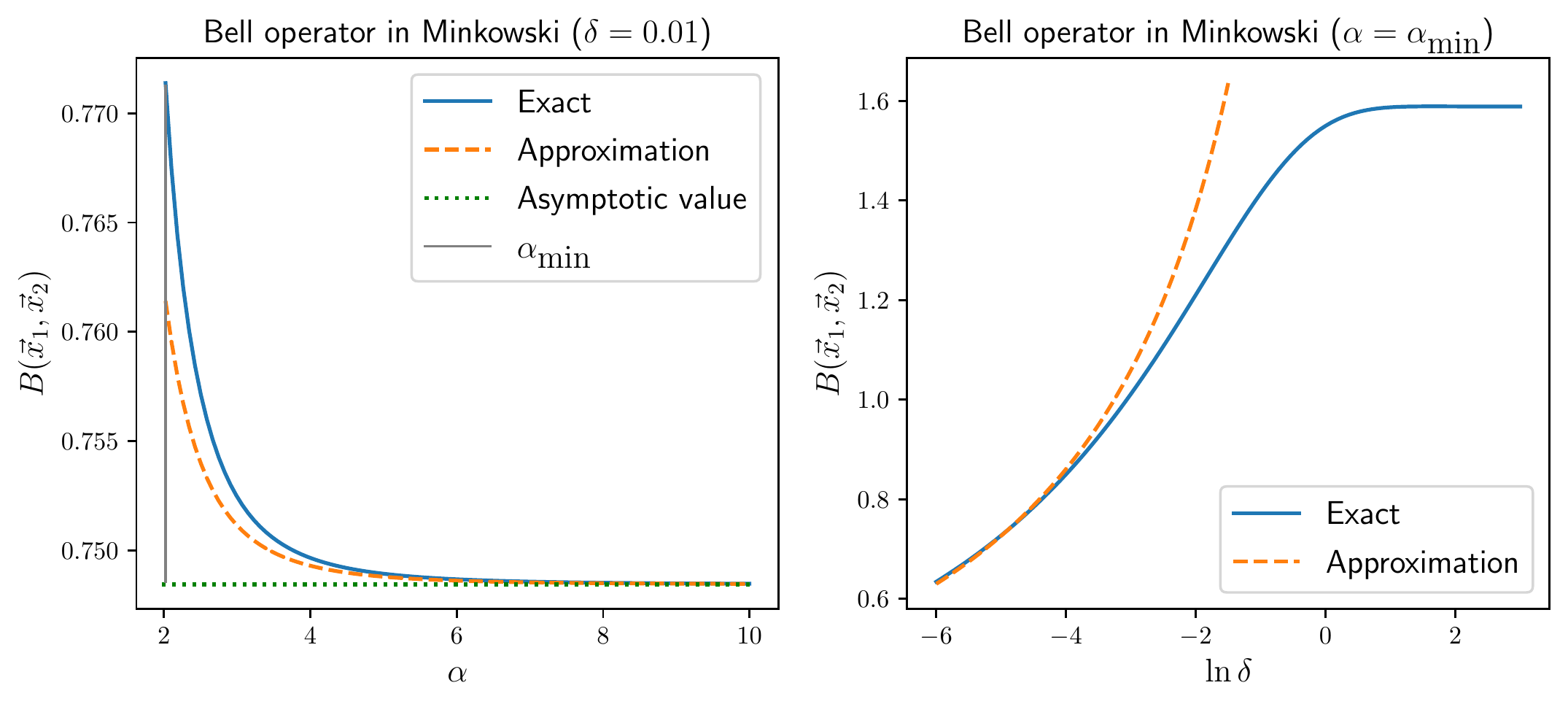}
    \caption{Expectation value of the GKMR Bell operator in the Minkowski vacuum. Left panel: $\delta = 0.01$ and $\alpha$ is varied from its minimum allowed value, $\alpha_\umin=2(1+\delta)$, to larger values. The blue solid line stands for the full result, the orange dashed line is obtained from the small-$\delta$, large-$\alpha$ approximation~\eqref{eq:Bell:Minkowski:Approx}, and the green dotted line corresponds to the asymptotic value at large $\alpha$. Right panel: $\alpha=\alpha_\umin$ (which maximises the expectation value of the Bell operator, see left panel) and $\delta$ is varied. The approximation~\eqref{eq:Bell:Minkowski:Approx}, displayed in orange, still provides a good fit to the full result at small $\delta$, even though $\alpha_{\mathrm{min}}$ is not so much larger than one. From these figures one concludes that Bell inequalities are never violated in this setup.}
    \label{fig:GKMR_Mink}
\end{figure}
By plugging those expressions into \Eq{eq:bell:Szz:Sxx}, one obtains an explicit formula for the expectation value of the Bell operator in terms of the two parameter $\alpha$ and $\beta$. The result is displayed in the left panel of \Fig{fig:GKMR_Mink} for $\delta=0.01$ and as a function of $\alpha$. One can see that $B(\vec{x}_1,\vec{x}_2)$ decreases with $\alpha$ and reaches an asymptotic value at large distances between the two patches. This can be understood by expanding the above formulas in the limit where $\delta$ is small but $\alpha$ is large, which gives rise to
\bea
\label{eq:Spin:Minkowski:Approx}
	\braket{\hat{\mathcal{S}}_x(\vec{x}_1) \hat{\mathcal{S}}_x(\vec{x}_2)} & \simeq \frac{8}{9\pi\alpha^2}(1+\delta)+\mathcal{O}\left(\frac{\delta^2}{\alpha^2},\frac{1}{\alpha^4}\right),\\
	\braket{\hat{\mathcal{S}}_z(\vec{x}_1) \hat{\mathcal{S}}_z(\vec{x}_2)} & \simeq\frac{4\pi^2}{9 \left\vert 1-2 \ln\frac{\delta}{2}\right\vert}\left[1+\frac{8}{81\alpha^4}+\mathcal{O}\left(\delta,\frac{1}{\alpha^6}\right)\right] .
\eea
Together with \Eq{eq:bell:Szz:Sxx}, this leads to
\bea
\label{eq:Bell:Minkowski:Approx}
B\simeq \frac{8\pi^2}{9\left\vert 1-2 \ln\frac{\delta}{2}\right\vert} \left\lbrace 1+\frac{2}{\alpha^4}\left[\frac{4}{81}+\left(\frac{1-2\ln\frac{\delta}{2}}{\pi^3}\right)^2\right]\right\rbrace\, ,
\eea 
which is displayed as the orange dashed line in \Fig{fig:GKMR_Mink}. One can check that, when $\alpha\gg 1$, it provides a good fit to the full result indeed. At large $\alpha$, $B$ reaches a constant that is given by the first term of \Eq{eq:Bell:Minkowski:Approx} and which is controlled by the purity of the state. It is displayed with the green dotted line. Note that the case $\delta=0$ is singular, and leads to $B\simeq 16/(9\pi\alpha^2)$ at large $\alpha$ (so the asymptotic value vanishes).

Since $B$ is maximal when $\alpha$ is minimal, in the right panel of \Fig{fig:GKMR_Mink} we set $\alpha$ to its minimal value and let $\delta$ vary, so as to optimise the expectation value of the Bell operator. One can see that $B$ increases with $\delta$,\footnote{When $\delta\to 0$, $\alpha_\umin\to 2$ and one has $\mathcal{L}_1=3(13-16\ln 2)/20$, $\mathcal{L}_3=3(\ln 2-1)/2$, $\mathcal{K}_1=9/4$ and $\mathcal{K}_3$ diverges logarithmically with $\delta$. This leads to $B\simeq 0.16$, which corresponds to the lower asymptotic value in the right panel of \Fig{fig:GKMR_Mink} that would be reached if the horizontal axis extended to large enough negative values.\label{footnote:Mink:delta:to:zero}} and reaches an asymptotic value at large $\delta$ of order $1.6$. This is therefore the largest value one can obtain in this setup, and since it is smaller than $2$, we conclude that GKMR Bell inequalities are never violated in flat space time.

\subsection{Discussion}
We thus conclude this section by reporting no real-space Bell-inequality violation with the GKMR pseudo-spin operators in the Minkowski vacuum state. This may be expected from the fact that no entangled particle are being created in that setup. However, as mentioned above, in \Refa{Martin:2021xml} it was pointed out that quantum discord does not vanish in the case under consideration. As a consequence, a non-vanishing discord does not seem to be necessarily related to Bell inequality violations, at least in this setup and with the spin operators considered here (see \Sec{sec:Other:Spin:Operators} for a generalisation to a larger class of spin operators). This is because, when performing measurements of a quantum field in real space, one effectively deals with a mixed system, for which the interpretation of quantum discord is less clear.

Let us finally note that, in \Refa{Martin:2021xml}, it was found that quantum discord decays as $\alpha^{-4}$ at large distances. The behaviour of quantum discord as a function of $\alpha$ is therefore the same as for the Bell expectation value, and in that sense, it may be seen as a useful tracer for identifying the configurations that are most likely to yield quantum effects. However, as we shall now see, this is not always true, since this behaviour similarity is lost in de-Sitter space times.
\section{De-Sitter space-time}
\label{sec:DeSitter}
Let us now turn our attention to de-Sitter space times, in order to address the case of primordial cosmological perturbations. A homogeneous and isotropic, spatially-flat universe is described by the Friedmann-Lema\^itre-Robertson-Walker metric
\bea
    \dd s^2 = a^2(\eta) \left(-\dd\eta^2 + \dd\vec{x}^2\right)\,,
\eea 
where $\vec{x}$ denotes comoving coordinates, $\eta$ is the conformal time (it is related to cosmic time $t$ via $\dd t = a \dd\eta$), and $a$ is the scale factor. During the inflationary epoch, the scale factor approximately grows exponentially with cosmic time, $a=\ee^{Ht}$ where $H=\dot{a}/a$ is the Hubble constant, which is referred to as de-Sitter geometry. In terms of conformal time, this leads to $a=-1/(H\eta)$, where $\eta$ varies from $-\infty$ to $0^-$. 

Cosmological perturbations can be described by a massless field $v$ evolving on this expanding background~\cite{Mukhanov:1981xt, Kodama:1985bj}. By setting its initial state in the Minkowski vacuum given in \Sec{sec:Minkowski}, one obtains the so-called Bunch-Davies vacuum state~\cite{Bunch:1978yq}, which is in excellent agreement with observations of the cosmic microwave background~\cite{Planck:2018jri}. The Fourier mode functions of the field $v$ and its conjugated momentum $p$ are thus given by
\bea
\label{eq:deSitter:mode:functions}
    v_{\vec{k}} = \frac{e^{-ik\eta}}{\sqrt{2k}} \left( 1- \frac{i}{k\eta} \right) \qquad \text{and} \qquad
    p_{\vec{k}} = v'_{\vec{k}} - \frac{a'}{a}v_{\vec{k}} = - i \sqrt{\frac{k}{2}} e^{-ik\eta}\, .
\eea 
Note that, strictly speaking, the curvature perturbation is ill-defined in de-Sitter backgrounds. We therefore assume that slow-roll deviations from the de-Sitter geometry are present, but that they are sufficiently small for \Eq{eq:deSitter:mode:functions} to provide a reliable approximation to the Mukhanov-Sasaki mode functions.
\subsection{Covariance matrix}
The reduced power spectra associated to the mode functions~\eqref{eq:deSitter:mode:functions} read
\bea
    \mathcal{P}_{vv} (k) = \frac{1+k^2 \eta^2}{4\pi^2 \eta^2}\, , \qquad
    \mathcal{P}_{pp} (k) = \frac{k^4}{4\pi^2}\, , \qquad
    \mathcal{P}_{vp} (k) = \frac{k^2}{4\pi^2 \eta}\,.
\eea
The covariance matrix is then obtained by plugging these expressions into \Eq{eq:covariance:matrix}, and this leads to
\bea
\label{eq:Covariance:deSitter}
    \gamma_{11} & = \frac{(HR)^2}{3\pi G(\delta)}   \left[ \mathcal{K}_{-1}(\beta,\delta) + \frac{1}{(HR)^2} \mathcal{K}_1(\beta,\delta)\right]\,,\\
    \gamma_{12} & = - \frac{HR}{3\pi G(\delta)} \mathcal{K}_1(\beta,\delta)\,,
    \qquad \gamma_{22} = \frac{\mathcal{K}_3(\beta,\delta)}{3\pi  G(\delta)}  \,,\\
    \gamma_{13} & = \frac{(HR)^2}{3\pi G(\delta)}   \left[ \mathcal{L}_{-1}(\alpha,\beta, \delta) + \frac{1}{(HR)^2} \mathcal{L}_1(\alpha, \beta, \delta)\right]\,,\\
    \gamma_{14} & = - \frac{HR}{3\pi G(\delta)}  \mathcal{L}_1(\alpha,\beta,\delta)\,, \qquad
    \gamma_{24} = \frac{\mathcal{L}_3(\alpha,\beta,\delta)}{3\pi  G(\delta)} \,.
\eea 
It depends on four parameters, namely $HR$, $\alpha$, $\beta$ and $\delta$. The parameters $\alpha$ and $\delta$ have already been introduced in \Sec{sec:Minkowski}, and we recall that they respectively correspond to the distance between the two patches in units of their radius, see \Eq{eq:alpha:def}, and to the smoothing parameter of the window function, see \Eq{eq:WindowFunction:Improved}. The parameter $HR$ corresponds to the ratio between the size of the patches and the Hubble radius, which is the typical distance that characterises the curvature of space time. The parameter $\beta$ is defined as the ratio between the size of the observed patches and the size of the entire observable universe, \ie the size of the region over which observations are performed,
\bea 
    \beta \equiv \frac{R}{R_{\mathrm{obs}}} <1 \, .
\eea 
Indeed, in practice, cosmological perturbations are measured as fluctuations away from an average configuration, where the average is computed as a mean value over a finite part of the universe, the size of which is denoted $R_{\mathrm{obs}}$. This implies that, in \Eq{eq:covariance:matrix}, the window function needs to be replaced according to $\widetilde{W}(kR/a) \to \widetilde{W}(kR/a) - \widetilde{W}(kR_{\mathrm{obs}}/a)$. This can be simply modelled by imposing an infra-red cutoff $k>\beta a/R$ in the integral of \Eq{eq:covariance:matrix}, see \Refa{Martin:2021qkg} for further details. The integrals $\mathcal{K}$ and $\mathcal{L}$ are thus defined in a similar way as in \Eq{eq:Integrals:K:L} but with $\beta$ as a lower bound, namely
\bea
\label{eq:Integrals:K:L:gen}
    \mathcal{K}_\mu(\beta,\delta) & = \int_\beta^\infty z^\mu \widetilde{W}^2(z) \dd z\, ,\\
    \mathcal{L}_\mu(\alpha,\beta,\delta) & = \int_\beta^\infty z^\mu \widetilde{W}^2(z) \left(z\right) \mathrm{sinc}(\alpha z) \dd z\, .
\eea
These integrals can be computed analytically in terms of the cosine integral function. The relevant formulas can be found in Appendix A of \Refa{Martin:2021qkg}, where a systematic expansion in the regime $\beta\ll 1$, $\delta\ll 1$ and $\alpha\gg 1$ is also performed. Let us finally note that, in the limit where $H=0$ (\ie static space time), \Eq{eq:Covariance:deSitter} boils down to the Minkowski formula~\eqref{eq:gamma:Minkowski}, as it should.
\
\subsection{Spin and Bell correlators}
\label{sec:deSitter:SpinAndBell}
\begin{figure}
    \centering
    \includegraphics[width=0.8\textwidth]{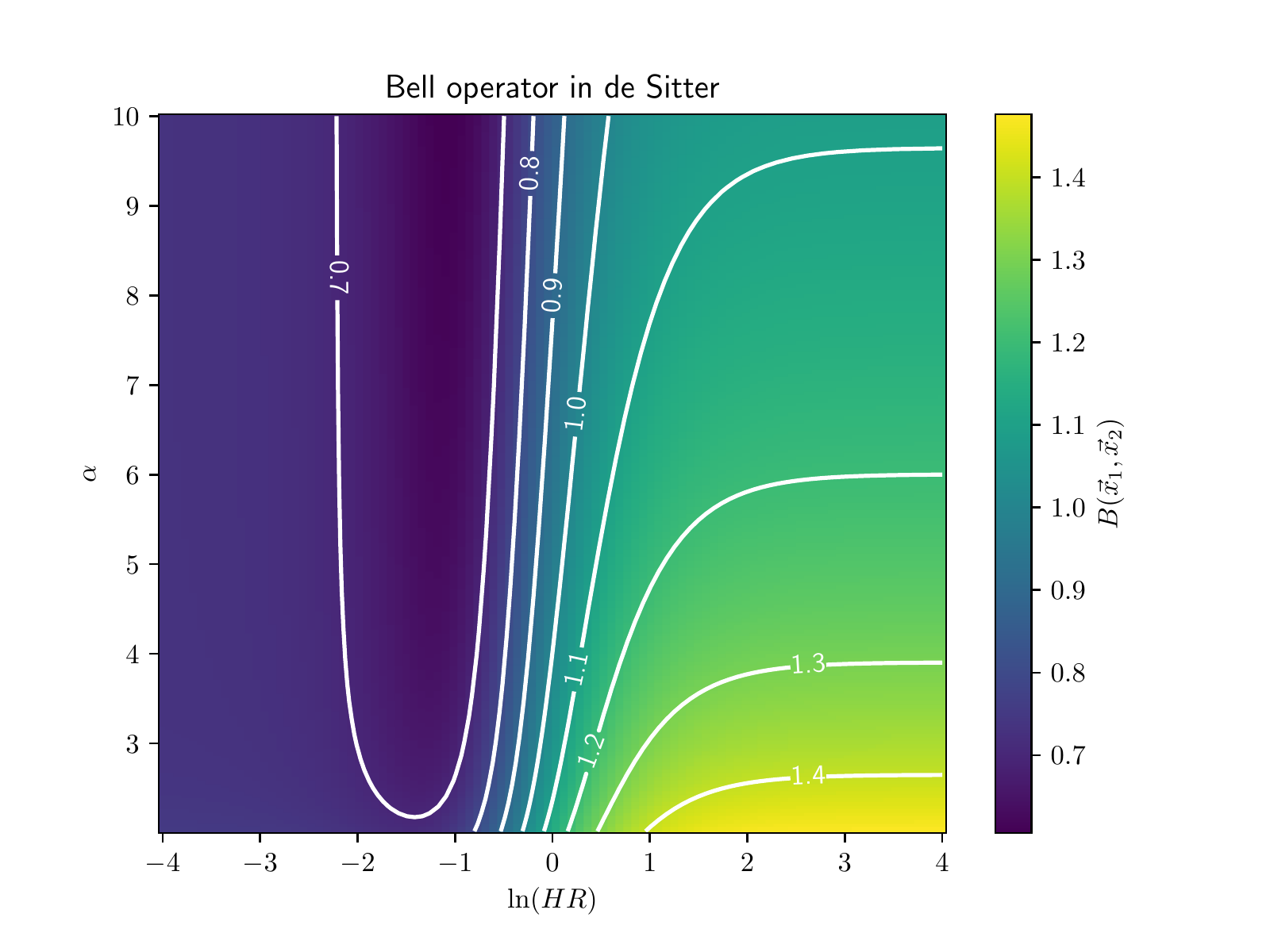}
    \caption{Expectation value of the GKMR Bell operator in the Bunch-Davies vacuum of the de-Sitter space-time, as a function of the parameters $\alpha=d/R$ and $HR$. Here $\alpha$ varies from its minimal value $\alpha_\umin = 2(1+\delta)$. The colour encodes the value of $B$, and a few contour lines are displayed in white. The UV and IR regulators (on which there is at most a logarithmic dependence) have been respectively set to $\delta = 0.01$ and $\beta=10^{-4}$ (see \App{app:Additional:Figures} for other slices in parameter space). Different behaviours are obtained depending on whether $HR<1$ or $HR>1$, \ie depending on the size of the measured patches with respect to the Hubble radius (see main text for further details).}
    \label{fig:GKMR_dS}
\end{figure}

By plugging the covariance matrix~\eqref{eq:Covariance:deSitter} into the formulas of \Sec{sec:SpinCorrelators}, the expectation value of the Bell operator can be computed, and the result is displayed in \Fig{fig:GKMR_dS}. In that figure, the UV and IR regulators have been respectively set to $\delta = 10^{-2}$ and $\beta=10^{-4}$. Since there is at most a logarithmic dependence of the result on these parameters, they do not play a crucial role, and \Fig{fig:GKMR_dS} rather shows how the result depends on $HR$ and $\alpha$ (see however \App{app:Additional:Figures} for figures displaying the dependence of $B$ on $\delta$ and $\beta$). 
One can see that two regimes need clearly to be distinguished, depending on whether $HR\ll 1$ (\ie the size of the patches is smaller than the Hubble radius) or $HR\gg 1$ (\ie the patches are larger than the Hubble radius). 

When $HR\ll 1$, the result seems to carry little dependence on $\alpha$, and  coincides with the values obtained in the left panel of \Fig{fig:GKMR_Mink} in the Minkowski vacuum. This is because, when $HR\ll 1/\alpha$, all distances involved in the problem (namely $R$ and $d$) are smaller than the Hubble radius, hence the setup is equivalent to a local Minkowski background. This is why the results of \Sec{sec:Minkowski} are recovered in this regime, which can be formally verified by expanding the above formulas in $HR$ and then in $\alpha^{-1}$, leading to
\bea 
&\braket{\hat{\mathcal{S}}_x(\vec{x}_1)\hat{\mathcal{S}}_x(\vec{x}_2)} \simeq
\frac{8}{9\pi \alpha^2} + 
(HR)^2 \left\lbrace \frac{8}{9\pi}
\left[\gamma_{\mathrm{E}}+\mathrm{ln} (\alpha\beta)-1\right] + \frac{32\left[5\gamma_{\mathrm{E}}-11 + 5 \ln (2\beta) \right]}{405 \alpha^2 \pi} 
 \right\rbrace \, ,\qquad \\
&\braket{\hat{\mathcal{S}}_z(\vec{x}_1)\hat{\mathcal{S}}_z(\vec{x}_2)}  \simeq 
\frac{4\pi^2}{9|1 - 2 \ln\frac{\delta}{2}|} + \frac{8 \pi^2}{81} (HR)^2 \\ &
\qquad \times \frac{1+ 2\gamma_{\mathrm{E}}(1+2\ln 2 ) - 5 \ln 2 + 4 \ln^2 2 + \ln \beta(2- 4 \ln \frac{\delta}{2}) + (7 - 4\gamma_{\mathrm{E}} - 4\ln 2) \ln \delta }{(1-2\ln\frac{\delta}{2})|(1-2\ln\frac{\delta}{2})|} .
\eea
In these expressions, $\gamma_{\mathrm{E}}$ is Euler's constant, and the result is further expanded in $\delta$ and $\beta$. One thus recovers \Eq{eq:Spin:Minkowski:Approx}, with corrections suppressed by $(HR)^2$.

\begin{figure}
    \centering
    \includegraphics[width=0.8\textwidth]{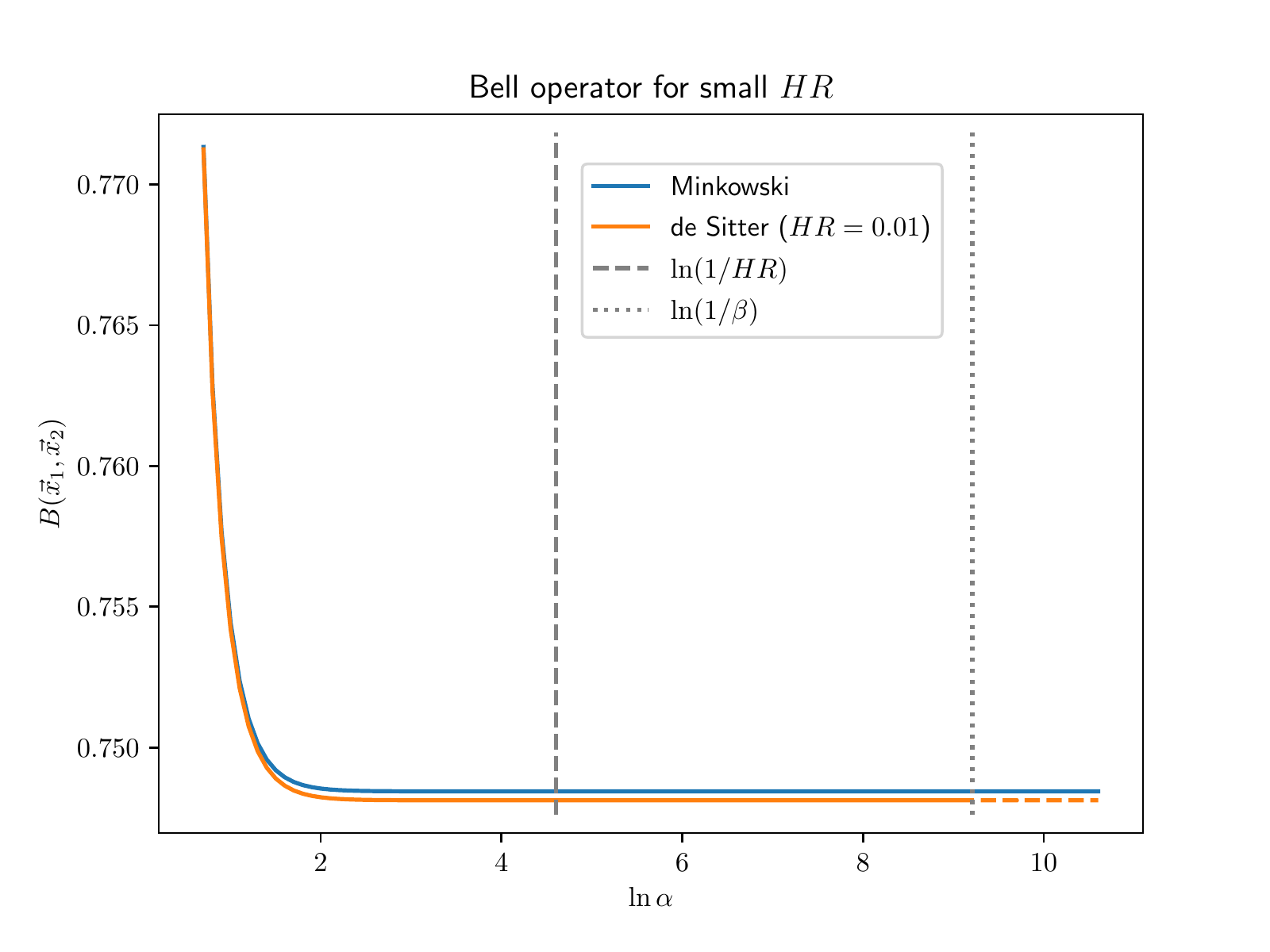}
    \caption{Expectation value of the GKMR Bell operator in the Bunch-Davies vacuum of the de-Sitter space time for $\beta = 10^{-4}$, $HR = 10^{-2}$, $\delta = 10^{-2}$, and as a function of the parameter $\alpha = d/R$. For $\alpha \ll 1/(HR)$, all distances involved in the problem are smaller than the Hubble radius hence the de-Sitter and Minkowski results coincide. Note that values of $\alpha > 1/\beta$ are not displayed since they would correspond to $d>R_{\mathrm{obs}}$.}
    \label{fig:dS_alphas}
\end{figure}
If $1/\alpha\ll HR\ll 1$, $R$ is smaller than the Hubble radius but not $d$, hence the flat space-time result may be modified a priori. However, this regime cannot be seen in \Fig{fig:GKMR_dS} since it does not display large-enough values of $\alpha$. This is why, in \Fig{fig:dS_alphas}, the expectation value of the Bell operator is shown as a function of $\alpha$, where $\beta$ and $\delta$ are fixed to the same value as in \Fig{fig:GKMR_dS}, and where we have set $HR=10^{-2}$. For comparison, the flat space-time result is also displayed. No strong deviation from the de-Sitter result can be observed, even when $d$ is larger than the Hubble radius. In any case, one can see that $B$ decreases with $\alpha$. Therefore, when $HR\ll 1$, $B$ is maximal in the Minkowski limit, where we have already shown that there is no Bell-inequality violation. 

When $HR\gg 1$, one can see in \Fig{fig:GKMR_dS} that an asymptotic value is also reached, which decreases with $\alpha$. This can be understood analytically by performing a large $HR$, large $\alpha$ expansion of the above formulas, which leads to
\bea
\label{eq:Larsson:approx:large:HR}
&\braket{\hat{\mathcal{S}}_x(\vec{x}_1)\hat{\mathcal{S}}_x(\vec{x}_2)} \simeq  \frac{2}{\pi} \arctan\left\lbrace { \frac{4 \left[1- \gamma_{\mathrm{E}} - \ln (\alpha \beta)\right]}{\sqrt{(3 + 4 \ln \frac{\alpha}{2})\left[11 - 8 \gamma_{\mathrm{E}} - 4 \ln (2 \alpha \beta^2)\right]}}  }\right\rbrace \, ,\\
&\braket{\hat{\mathcal{S}}_z(\vec{x}_1)\hat{\mathcal{S}}_z(\vec{x}_2)} \simeq  \frac{2\pi^2}{(HR)^2} \Bigg\lbrace \bigg[4\gamma_{\mathrm{E}}(1 + 2 \ln 2) -1 - 9 \ln 2 + 4\ln^2 2   + 2\ln(\alpha \beta^2) \left(1 - 2 \ln \frac{\delta}{2}\right)\quad\quad\\
&\quad + \ln \delta\left(11 - 8\gamma_{\mathrm{E}} - 4 \ln 2\right)\bigg]\left[3 - \ln 2 + 4 \ln^2 2  
- \ln \alpha \left(2 - 4 \ln \frac{\delta}{2}\right)+  \left(3 - 4\ln 2\right) \ln \delta
\right] \Bigg\rbrace^{-1/2}\, .
\eea

The conclusion of this analysis is that the configuration leading to the maximum expectation value for the Bell operator is the one where $HR$ is large and $\alpha$ is close to its minimum value. This corresponds to the situation where the coarse-graining scale $R$ is large compared to the Hubble radius, and the two patches are almost adjacent. In \App{app:Additional:Figures}, we further show that decreasing $\delta$ and $\beta$ make $B$ larger. This allows us to derive an upper bound on $B$ as follows. Formally, when $\beta\to 0$, the integrals $\mathcal{K}_{-1}$ and $\mathcal{L}_{-1}$ logarithmically diverge, and in the limit $\delta\to 0$, $\mathcal{K}_{3}$ logarithmically diverges too. This behaviour is such that  $\mathfrak{p}=\braket{\hat{\mathcal{S}}_z(\vec{x}_1) \hat{\mathcal{S}}_z(\vec{x}_2)}\propto 1/\ln(\beta\delta)\to 0$ in this limit, and such that the argument of the $\arctan$ function in \Eq{eq:SxSx} goes to a finite constant (that only depends on $\alpha$ if one further lets $HR\to \infty)$. This proves that $\braket{\hat{\mathcal{S}}_z(\vec{x}_1) \hat{\mathcal{S}}_z(\vec{x}_2)}<1$ in this limit, hence $B<2$, see \Eq{eq:bell:Szz:Sxx}, which therefore applies to the whole parameter space.

%There is a clear transition at $HR \sim 1$ between the sub-horizon and super-horizon regimes. For $HR \ll 1$, there is little dependence of $\braket{B(\vec{x},\vec{y})}$ on $\alpha$, while for $HR \gg 1$ it converges to a fixed value that does decrease with $\alpha$. Recall that the Bunch-Davies vacuum is obtained by imposing that the mode functions of the scalar field reduce to those of Minkowski vacuum, i.e. plane waves, in the sub-Hubble limit. Even though the value of $\braket{B(\vec{x},\vec{y})}$ for the Minkowski vacuum does depend on $\alpha$, its largest contribution comes from the purity, which has a zeroth order term in $\alpha$, unless the limit $\delta \rightarrow 0$ is taken. More precisely, the purity is roughly negligible as long as $\alpha^2 \ll \frac{8}{\pi} |\log \delta|$.

%The largest value of $\braket{B(\vec{x},\vec{y})}$ is achieved for small $\alpha$ in the large $HR$ regime. However, it does not seem to exceed the threshold $\braket{B(\vec{x},\vec{y})} = 2$ imposed by local realism. Even though the precise value depends on the parameters $\beta$ and $\delta$, the overall physical picture does not.
%
\subsection{Discussion}
The fact that no Bell-inequality violation is found in de Sitter is non trivial. This is because, due to space-time curvature, entangled pairs of particles with opposite Fourier momenta are massively produced on super-Hubble scales. In Fourier space, this leads to large quantum squeezing, to a large quantum discord~\cite{Martin:2015qta, Martin:2021znx}, as well as Bell-inequality violations (both with the GKMR operators and with other pseudo-spin operators, see \Refs{Martin:2016tbd, Martin:2017zxs}). The reason why this does not translate into Bell-inequality violations in real space is that, as mentioned above, when going to real space, one has to deal with effectively mixed states. This was also seen at the level of quantum discord in \Refa{Martin:2021qkg}, which was found to be much smaller in real space than in Fourier space. However, a non-vanishing quantum discord was still obtained, which suggests that quantum discord may not always provide a direct quantum criterion in the context of mixed states.

\begin{figure}
    \centering
    \includegraphics[width=0.80\textwidth]{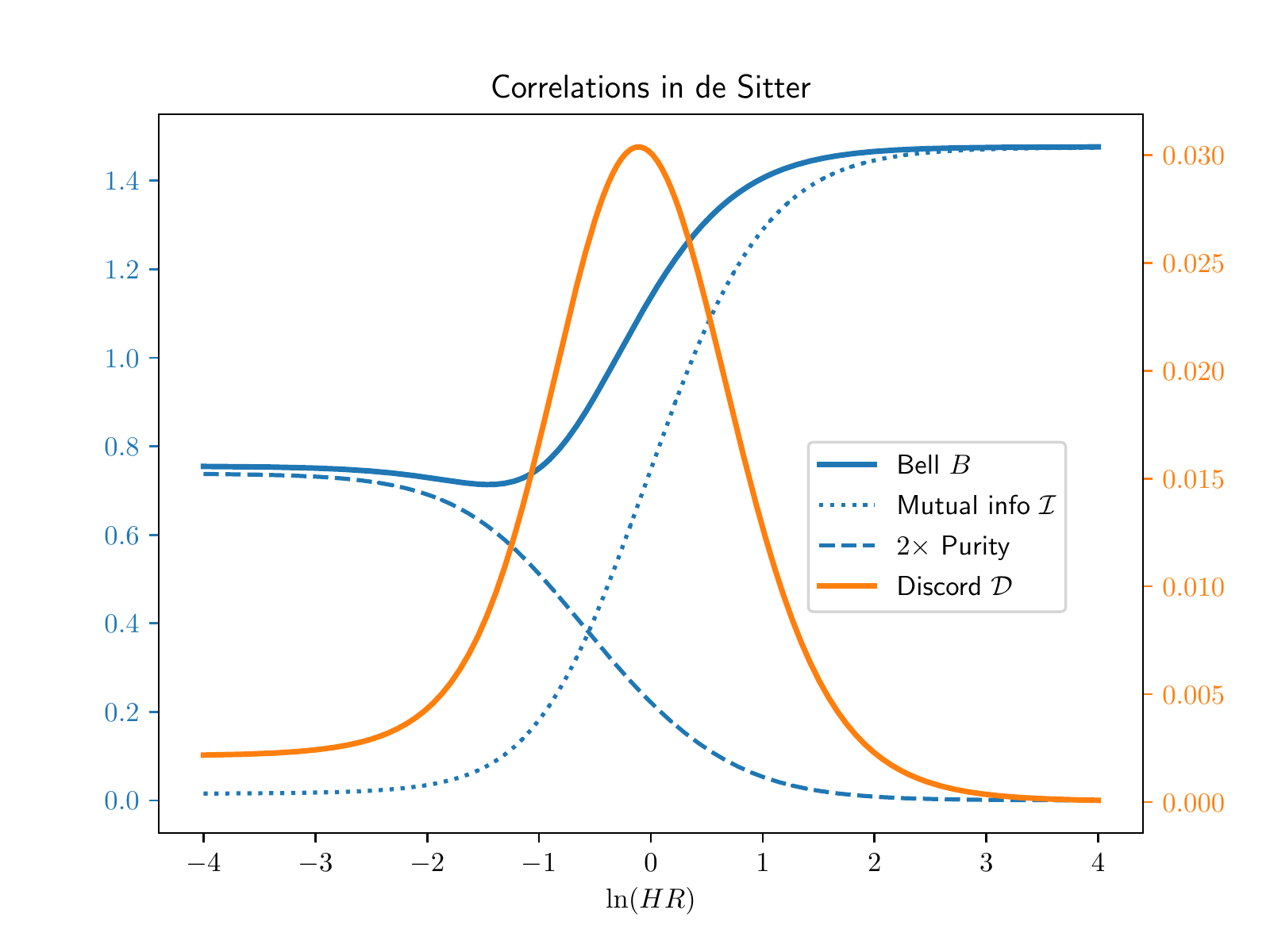}
    \caption{Expectation value of the GKMR Bell operator $B$ in the de-Sitter space time, as a function of $HR$ and for $\alpha = \alpha_\umin = 2(1+\delta)$, $\beta = 10^{-4}$ and $\delta = 10^{-2}$. For comparison, we also display (two times) the state purity $\mathfrak{p}$, to which $B$ asymptotes in the small-$HR$ limit, as well as mutual information $\mathcal{I}$ and quantum discord $\mathcal{D}$, the latter being labelled with the rightmost vertical axis.}
    \label{fig:GKMR_purity_dS}
\end{figure}
In order to better discuss the relationship between quantum discord and Bell inequalities, in \Fig{fig:GKMR_purity_dS} we display both quantities as a function of $HR$, for fixed values of $\alpha$, $\beta$ and $\delta$. One recovers that $B$ approaches a constant when $HR\ll 1$ (which corresponds to the Minkowski limit) as well as when $HR\gg 1$, with a small dip between the two regimes, the later plateau being slightly higher than the former.
In contrast, quantum discord vanishes in these two limits, and is maximal when $HR$ is of order one. Therefore, contrary to the flat space-time case discussed in \Sec{sec:Minkowski}, discord cannot be used to identify the setup configuration that maximises our ability to detect quantum features. 

For comparison, the state purity introduced in \Eq{eq:purity:def} is also shown in \Fig{fig:GKMR_purity_dS}. In the limit $HR\ll 1$, it shares the same behaviour as $B/2$, as already explained in \Sec{sec:Minkowski}. In this regime, the more mixed the quantum state is, the smaller the value of $B$ is, which is intuitive. However, in the large $HR$ limit, the behaviour of the state purity and of the Bell expectation value are opposite: $\mathfrak{p}$ decreases with $HR$ while $B$ increases. This can be understood as follows. Since more entangled particle are created at large scales, the field becomes more correlated in real space as $R$ increases, as can be seen at the level of the mutual information $\mathcal{I}$, which only increases with $HR$. It explains why the state purity decreases (one traces over regions of space to which the system is more and more entangled).
Both quantum discord and the Bell operator are driven by a compromise between the amount of quantum entanglement (measured by mutual information $\mathcal{I}$) and the state purity $\mathfrak{p}$. But since these two quantities evolve in opposite ways, how the  trade-off is settled is a priori not trivial, and it happens to be settled in different ways for $\mathcal{D}$ and $\mathcal{B}$. 
%As a consequence, discord does not seem to provide a reliable tracer of quantum features in the present situation.
%
\section{Other pseudo-spin operators}
\label{sec:Other:Spin:Operators}
So far we have shown that real-space Bell inequalities cannot be violated with the GKMR pseudo-spin operators, both in flat space-time and in de-Sitter cosmologies. However, one may argue that this result is restricted to a specific class of pseudo-spin operators, and that violations may be obtained by considering other operators. Unfortunately, it is not possible to verify (at least with the present approach) all possible pseudo-spin operators, given that there may exist an infinite number of them and that only a few explicit constructions are known~\cite{2009JPhA...42B5309D, Martin:2017zxs}. However, a basic sanity check is to study another set of pseudo-spin operators, which is the goal of this section. In practice, we consider the Larsson spin-operators~\cite{2004PhRvA..70b2102L}, which is in fact an infinite, one-parameter family of spin operators.
\subsection{Larsson pseudo-spin operators}
The idea of the Larsson pseudo-spin operators is to split the real axis describing the scalar field value into intervals of size $\ell$, where $\ell$ can be freely chosen by the observer. One then introduces~\cite{2004PhRvA..70b2102L}
\bea
\label{eq:Sz:Larsson}
	\hat{S}_x^\ell (\vec{x}) & = \sum_{n=-\infty}^{\infty}  \int_{2n\ell}^{(2n+1)\ell} \dd \tilde{\phi}_R(\vec{x}) \left[ \ket{\tilde{\phi}_R(\vec{x})+\ell}{\bra{\tilde{\phi}_R(\vec{x})}} + \ket{\tilde{\phi}_R(\vec{x})}{\bra{\tilde{\phi}_R(\vec{x}) + \ell}} \right],\\
	\hat{S}_y^\ell (\vec{x}) & = - i \sum_{n=-\infty}^{\infty}  \int_{2n\ell}^{(2n+1)\ell} \dd \tilde{\phi}_R(\vec{x}) \left[  \ket{\tilde{\phi}_R(\vec{x})+\ell}{\bra{\tilde{\phi}_R(\vec{x})}} - \ket{\tilde{\phi}_R(\vec{x})}{\bra{\tilde{\phi}_R(\vec{x}) + \ell}} \right],\\
	\hat{S}_z^\ell (\vec{x}) & = \sum_{n=-\infty}^{\infty} (-1)^n \int_{n\ell}^{(n+1)\ell} \dd\tilde{\phi}_R(\vec{x}) \ket{\tilde{\phi}_R(\vec{x})}{\bra{\tilde{\phi}_R(\vec{x})}} .
\eea 
One can check that these operators are indeed pseudo-spin operators, \ie they satisfy the relations given below \Eq{eq:GKMR:def}. Their Wigner-Weyl transforms are given by~\cite{Martin:2017zxs}
\bea
\label{eq:Weyl:Larsson}
  W_{\hat{S}_x^\ell (\vec{x})} & =\sum_{n=-\infty}^\infty 2 \cos\left[\tilde{\pi}_R(\vec{x}) \ell\right] \left\lbrace \theta\left[\tilde{\phi}_R(\vec{x}) - \frac{\ell}{2} -2n\ell\right] - \theta\left[\tilde{\phi}_R(\vec{x}) - \frac{\ell}{2} -(2n+1)\ell\right] \right\rbrace,\\
  W_{\hat{S}_y^\ell (\vec{x})} & = \sum_{n=-\infty}^\infty 2 \sin\left[\tilde{\phi}_R(\vec{x}) \ell\right] \left\lbrace\theta\left[\tilde{\phi}_R(\vec{x}) - \frac{\ell}{2} -2n\ell\right] - \theta\left[\tilde{\phi}_R(\vec{x}) - \frac{\ell}{2} -(2n+1)\ell\right] \right\rbrace,\\
  W_{\hat{S}_z^\ell (\vec{x})} & = \sum_{n=-\infty}^\infty (-1)^n \left\lbrace \theta\left[\tilde{\phi}_R(\vec{x}) - n\ell\right] - \theta\left[\tilde{\phi}_R(\vec{x}) - (n+1)\ell\right]\right\rbrace ,
\eea
where $\theta$ is the Heaviside function. By plugging these expressions into \Eq{eq:mean:weyl}, one obtains explicit expressions for the spin correlators in terms of double infinite sums of integrals involving the error function, which are given in \App{sec:Larsson:Spin:Correlators}.

These expressions can be evaluated numerically. Before displaying the result, it is worth mentioning that further analytical insight can be gained by expanding those formulas in the limits $\ell\ll 1$ and $\ell\gg 1$. Those expansions are carried out in \App{sec:Larsson:Spin:Correlators} and below we only give the final result. 

When $\ell\ll 1$, one obtains
\bea
\label{eq:Larsson:low:ell}
\braket{\hat{S}^\ell_x(\vec{x}_1) \hat{S}^\ell_x(\vec{x}_2)} &\simeq \frac{1}{4} \sum_{\boldmathsymbol{\epsilon}=(\pm 1,\pm 1)^{\mathrm{T}}} e^{ \frac{\ell^2}{2}
\left\lbrace \frac{a_{11} \tilde{a}_2^2 + a_{22} \tilde{a}_1^2 - 2 a_{12} \tilde{a}_1 \tilde{a}_2 }{a_{11} a_{22} - a_{12}^2} +  \boldmathsymbol{\epsilon}^T\cdot [(\gamma^{-1})^{\pi\pi}]\cdot\boldmathsymbol{\epsilon} \right\rbrace } \underset{\ell\to 0}{\longrightarrow} 1\, ,\\
\braket{\hat{S}^\ell_z(\vec{x_1}) \hat{S}^\ell_z(\vec{x}_2)} &\simeq e^{-\frac{\pi^2 (a_{11} + a_{22} - 2 a_{12})}{2\ell^2 (a_{11} a_{22} - a_{12}^2)}}\underset{\ell\to 0}{\longrightarrow} 0\, ,
\eea
where $\tilde{a}_1$ and $\tilde{a}_2$ are functions of the entries of the covariance matrix given in \App{sec:Larsson:Spin:Correlators}. This shows that $B^\ell \rightarrow 2$ as $\ell \rightarrow 0$, hence there is no Bell-inequality violation in this regime. Let us note that, when $\ell$ decreases, the size of the $\ell$-intervals in \Eq{eq:Sz:Larsson} decreases, hence numerically one has to include more terms before truncating the sum. This is why the small-$\ell$ regime is numerically challenging, and there is always a minimum value of $\ell$ below which the computation cannot be performed, given finite numerical capacities. For this reason, having an analytical control on the small-$\ell$ regime is necessary to make sure there is no blind spot in the analysis.

When $\ell\gg 1$, one finds that 
\bea
\label{eq:larsson:SxSx:large:ell}
\braket{\hat{S}^\ell_x(\vec{x}_1) \hat{S}^\ell_x(\vec{x}_2)}
\underset{\ell\to \infty}{\longrightarrow} 0\, ,
\eea 
while $\hat{S}^\ell_z(\vec{x})$ approaches the $\hat{\mathcal{S}}_x$ component of the GKMR operator,
\bea
\label{eq:larsson:SzSz:large:ell}
\hat{S}_z^\ell (\vec{x}) & \underset{\ell\to \infty}{\longrightarrow}\hat{\mathcal{S}}_x^\ell (\vec{x}) .
\eea 
Its two-point function is thus given by \Eq{eq:SxSx} in this limit. Considering \Eq{eq:bell:Szz:Sxx}, this shows that $B^\ell$ is smaller than the GKMR result in the limit $\ell\to\infty$, hence no Bell-inequality violation can be obtained in this regime either. 

In between those two regimes, as mentioned above, one has to resort to numerical computations.
\subsection{Flat space-time}
\begin{figure}
    \centering
    \includegraphics[width=0.8\textwidth]{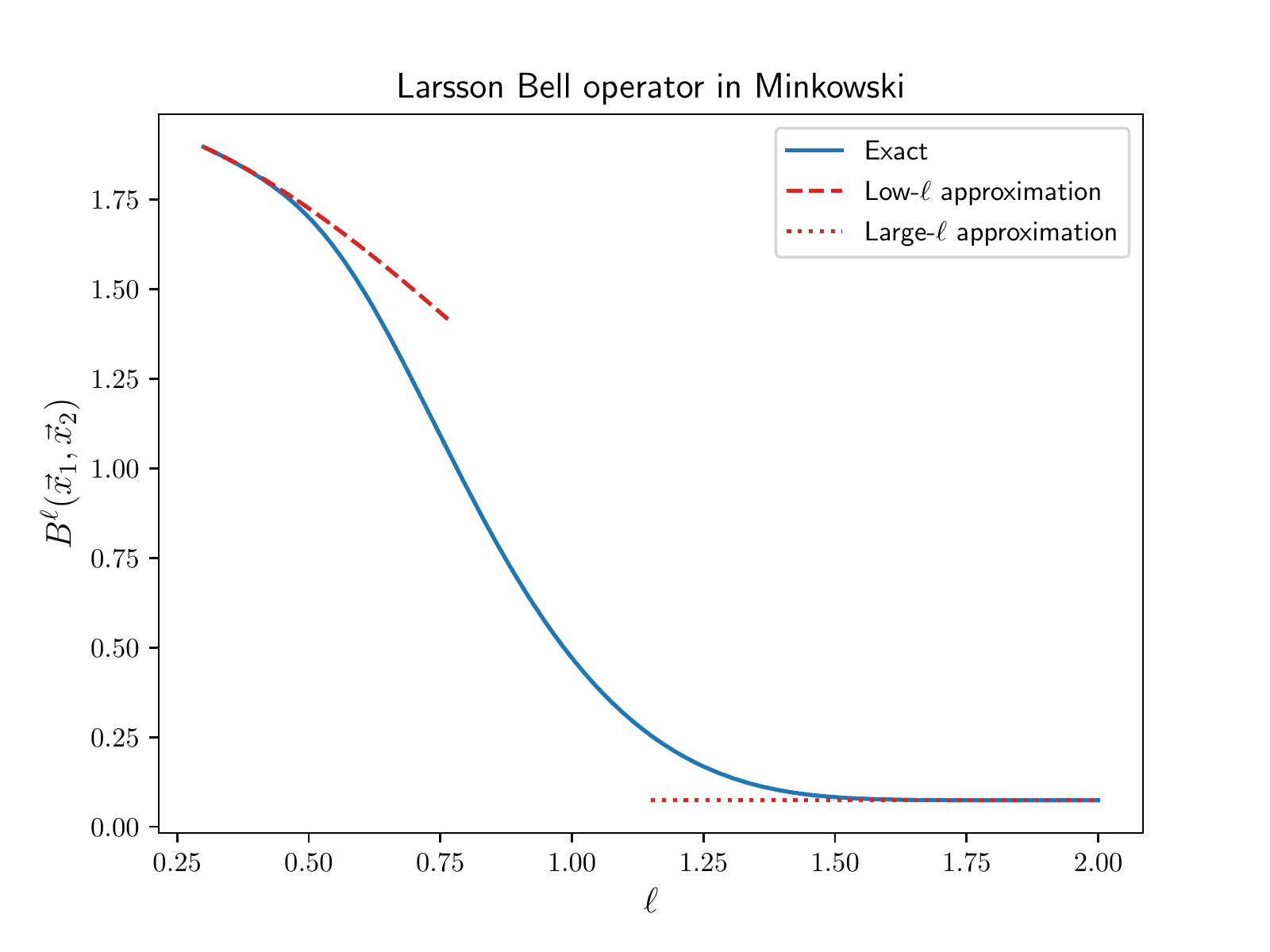}
    \caption{Expectation value of the Larsson Bell operator in the Minkowski vacuum, as a function of $\ell$, for $\alpha = 3$ and $\delta = 0.1$. The blue solid line corresponds to the full result, the red dashed line to the low-$\ell$ approximation~\eqref{eq:Larsson:low:ell}, and the red dotted line to the large-$\ell$ limit~\eqref{eq:larsson:SxSx:large:ell} and~\eqref{eq:larsson:SzSz:large:ell}.}
    \label{fig:Mink_ell}
\end{figure}
In \Fig{fig:Mink_ell}, we display the expectation value of the Larsson Bell operator in the Minkowski vacuum, as a function of $\ell$, for $\alpha = 3$ and $\delta = 0.1$. One can check that the small-$\ell$ and the large-$\ell$ approximations derived above provide good fits to the full result in their respective domains of validity. In between, we find that $B^\ell$ is a decreasing function of $\ell$, and this behaviour is observed for any value of $\delta$ and $\alpha$. As a consequence, one has $B^\ell<B^{\ell\to 0}=2$, hence there is no Bell-inequality violation in flat space time.
\subsection{De-Sitter space-time}
\begin{figure}
    \centering
    \includegraphics[width=\textwidth]{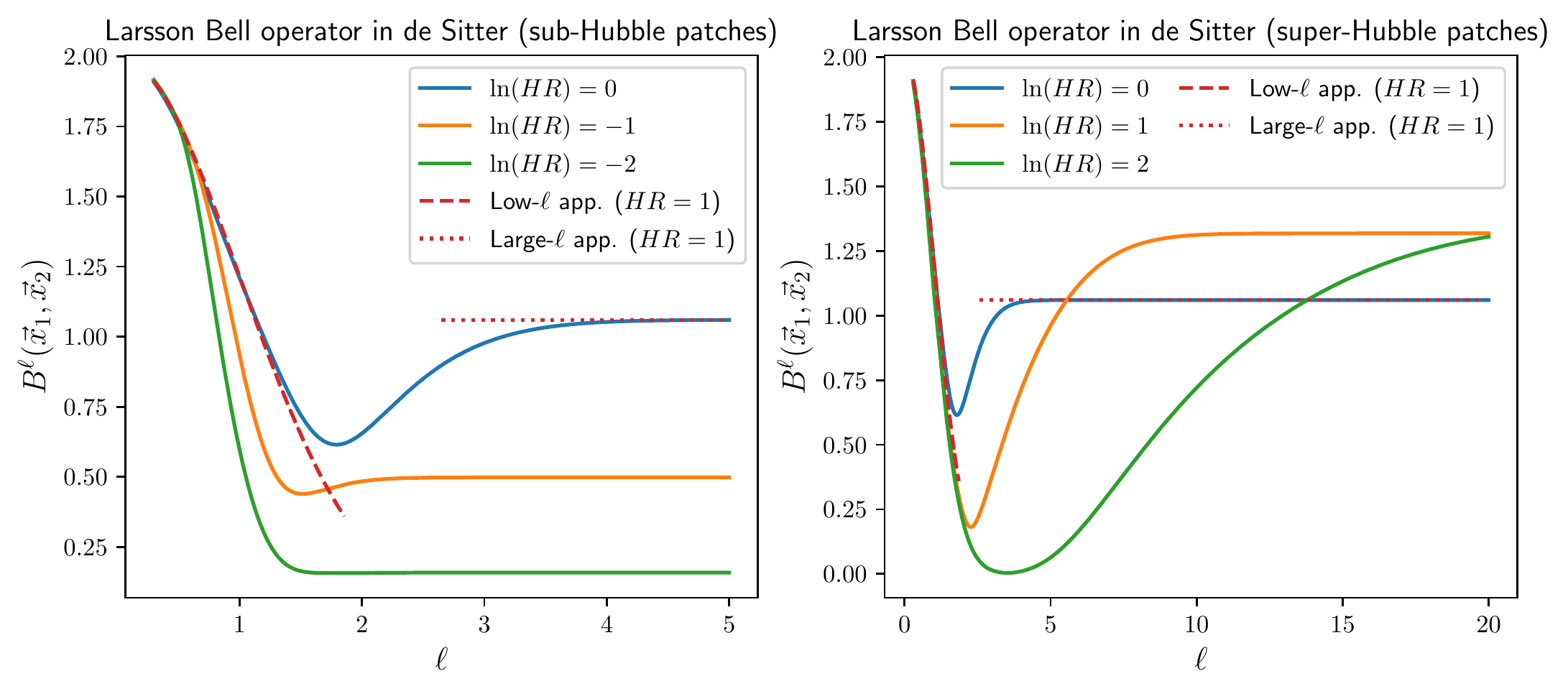}
    \caption{Expectation value of the Larsson Bell operator in the de-Sitter space-time, as a function of $\ell$, for $\alpha = 3$ and $\delta = 0.1$ and $\beta = 10^{-4}$. A few sub-Hubble values of $HR$ are shown in the left panel, and a few super-Hubble values in the right panel. The red dashed line corresponds to the low-$\ell$ approximation~\eqref{eq:Larsson:low:ell}, and the red dotted line to the large-$\ell$ limit~\eqref{eq:larsson:SxSx:large:ell} and~\eqref{eq:larsson:SzSz:large:ell}, both in the case $HR=1$.}
    \label{fig:dS_ell}
\end{figure}
Similarly, the expectation value for the Larsson Bell operator is displayed as a function of $\ell$ in the de-Sitter Bunch-Davies vacuum state in \Fig{fig:dS_ell}, for $\delta = 0.1$, $\beta = 10^{-4}$, $\alpha = 3$ and for a few values of $HR$. One can check that the small-$\ell$ and the large-$\ell$ approximations derived above still provide good fits to the full result. In between, $B^\ell$ goes through a local minimum at intermediate values of $\ell$, hence it is always smaller than $2$. This same behaviour is observed with other values for the parameters  $\delta$, $\beta$ and $\alpha$, which allows us to conclude that no Bell-inequality violation can be obtained with the Larsson operators. 
\section{Conclusion}
\label{sec:Conclusion}
In this work, we have shown how real-space Bell operators can be constructed for a quantum field, and how their expectation value can be obtained from the power spectra of the field if it is placed in a Gaussian state. We have then applied our formalism to the case of cosmological perturbations, during the primordial inflationary phase where space-time geometry is close to a de-Sitter background. We have found that no Bell-inequality violations can be reported.

This result had been paved by a series of previous works, starting with the computation of quantum discord in Fourier space for two-mode squeezed states~\cite{Lim:2014uea,Martin:2015qta}. This revealed that the creation of pairs of particles with opposite Fourier momenta in de-Sitter geometries is associated with the production of a large quantum discord, \ie with the presence of genuine quantum correlations. Fourier-space Bell operators were then constructed in \Refs{Martin:2016tbd, Martin:2017zxs}, confirming that Bell inequalities between opposite Fourier modes can indeed be violated, and hinting towards the presence of quantum features in primordial fluctuations. However, one of the assumptions on which the Bell's theorem rests is that of locality (\ie spacelike-separated events cannot influence each other), so Bell's inequalities should rather be tested in real space. 

This is why the presence of entanglement entropy and quantum discord between two distinct regions of space was then studied in \Refs{Espinosa-Portales:2019peb, Espinosa-Portales:2020pjp, Martin:2021xml, Martin:2021qkg}, where it was found that, although they remain non vanishing, their typical values are greatly reduced compared to the Fourier-space setup, casting some doubt on the ability of cosmological structures to display quantum correlations. The reason for the reduction of quantum discord when going from Fourier to real space is that, when considering two given regions of space, one implicitly traces the system over the field configuration in every other region of space, thus creating an effectively mixed state. This effect was dubbed ``effective decoherence''.

It thus remained to investigate how it may also affect Bell inequalities, which is the topic of the present work. Using the formalism of \Refa{Martin:2021xml}, we have shown how the two-point functions of pseudo-spin operators can be computed in Gaussian fields. 
We have first applied our framework to scalar fields placed in the vacuum state of the Minkowski space-time, and found that Bell inequalities are not violated in flat backgrounds. This result may seem a priori obvious (since no quantum particle is being created, there is no support for potential quantum correlations), but it is still instructive since it was found in \Refa{Martin:2021xml} that real-space quantum discord does not vanish in that setup. This illustrates that the interpretation of quantum discord for mixed states (with which we have to deal because of ``effective decoherence'') is indeed subtle.

We have then considered the Bunch-Davies vacuum state of de-Sitter space-times, in which cosmological perturbations are placed during inflation. In that case too, in spite of the efficient particle creation process operating on super-Hubble scales, we found no violation of the Bell inequalities. This suggests that the ``effective decoherence'' mechanism acts as a quantumness censor, at least in the setup considered here. 

Let us also recall that real-space quantum discord does not vanish in this state. Moreover, the configuration where it is maximal (namely when the two patches have size comparable to the Hubble radius and are almost adjacent~\cite{Martin:2021qkg}) does not coincide with the one where the expectation value of the Bell operator is maximal (namely when the patches are large compared to the Hubble radius and are almost adjacent). This illustrates again the subtleties of quantum discord for mixed states.\\

It is worth mentioning a few possible directions along which this research program could be carried on. First, even though we have generalised our finding to another family of spin operators in \Sec{sec:Other:Spin:Operators}, we have not tested \emph{all} possible Bell operators (only a few explicit constructions are known). Therefore, strictly speaking, we cannot claim that real-space Bell inequalities can never be violated in de Sitter, and it would be interesting to derive a generic mathematical argument (or expose a counter-example).

Second, there are other classes of Bell inequalities, which rely on measuring the system at different times. Those are the temporal Bell inequalities~\cite{1993PhRvL..71.3235P, 2004quant.ph..2127B, 2010NJPh...12h3055F}, the Legget-Garg inequalities~\cite{Leggett:1985zz}, and the bipartite temporal Bell inequality~\cite{1986NYASA.480..263B}. The last two were shown to be violated by cosmological perturbations in Fourier space in \Refs{Martin:2016nrr} and~\cite{Ando:2020kdz} respectively, and their investigation in real space remains to be carried out.

Third, it would be interesting to investigate correlations and entanglement between more than two spheres. This may effectively reduce the size of the traced-out regions, hence the importance of the ``effective decoherence'' effect. This may also require to account for non-Gaussianities, which have been shown to be relevant for the search of primordial quantum signals~\cite{Green:2020whw,Howl:2020isj}.  

Finally, on top of the ``effective decoherence'' effect, a ``physical decoherence'' mechanism may take place in the context of cosmology, arising from the fact that the scalar field describing cosmological adiabatic perturbations usually couples to other, unobserved degrees of freedom (additional fields, unobserved scales, etc.). This may come as an additional quantumness censor and would have to be studied, if one of the previously-mentioned directions turns out successful. Recently, in \Refa{Martin:2021znx}, it was found that there exists a wide region in parameter space where the Fourier-space quantum discord is unaffected by environmental effects even where they make the state of the system fully decohere. Generalising this calculation to real-space setups would further test the relationship between quantum discord, Bell-inequalities violation, and the detectability of quantum features in cosmological fields. 

\section*{Acknowledgements}
It is a pleasure to thank Juan Garc\'ia-Bellido for interesting discussions. The work of L.E.P. is funded by a fellowship from ``La Caixa" Foundation (ID 100010434) with fellowship code LCF/BQ/IN18/11660041 and the European Union Horizon 2020 research and innovation programme under the Marie Sklodowska-Curie grant agreement No. 713673. L.E.P. thanks APC Paris for the hospitality during his stay and acknowledges support by the PIF-UAM and UAM-Santander programs. This work is partially supported by the Spanish Research Agency (Agencia Estatal de Investigación) through the Grant IFT Centro de Excelencia Severo Ochoa No CEX2020-001007-S, funded by MCIN/AEI/10.13039/501100011033; as well as by Grant PGC2018-094773-B-C32 funded by MCIN/AEI/10.13039/501100011033 and by ERDF A way of making Europe.
\appendix

\section{GKMR spin correlators}
\label{sec:GKMR:Spin:Correlators}

In this appendix, we compute the three spin correlators $\braket{\hat{\mathcal{S}}_x(\vec{x}_1) \hat{\mathcal{S}}_x(\vec{x}_2)}$, $\braket{\hat{\mathcal{S}}_x(\vec{x}_1) \hat{\mathcal{S}}_z(\vec{x}_2)}$ and $\braket{\hat{\mathcal{S}}_z(\vec{x}_1) \hat{\mathcal{S}}_z(\vec{x}_2)}$ given in \Sec{sec:SpinCorrelators}. Plugging \Eq{eq:Wigner:Gauss} and \Eq{eq:Weyl:Transform} into \Eq{eq:mean:weyl}, one first has
\bea 
\label{eq:SxSx:interm0}
	\braket{ \hat{\mathcal{S}}_x(\vec{x}_1) \hat{\mathcal{S}}_x(\vec{x}_2)} & = \frac{1}{(2\pi)^2 \sqrt{\det \boldmathsymbol{\gamma}}}\int \dd\boldmathsymbol{q}\  \textrm{sign}(q_1) \textrm{sign}(q_3)\ee^{-\frac{1}{2}\boldmathsymbol{q}^{\mathrm{T}} \boldmathsymbol{\gamma}^{-1}\boldmathsymbol{q}}\, .
\eea
Since the field and momentum coordinates play different roles in this integral, let us first re-arrange the entries of the $\boldmathsymbol{q}$ vector such that the field coordinates appear first, and then the momentum coordinates:
\bea
\bar{\boldmathsymbol{q}} = 
\begin{pmatrix}
\tilde{\phi}_R(\vec{x}_1)\\
\tilde{\phi}_R(\vec{x}_2)\\
\tilde{\pi}_R(\vec{x}_1)\\
\tilde{\pi}_R(\vec{x}_2)
\end{pmatrix} = 
\underbrace{
\begin{pmatrix}
1 & 0 & 0 & 0\\
0 & 0 & 1 &0\\
0 & 1 & 0 & 0\\
0 & 0& 0& 1
\end{pmatrix}}_{\boldmathsymbol{P}}
\boldmathsymbol{q}\, ,
\eea
which defines the permutation matrix $\boldmathsymbol{P}$. In this new basis, the matrix $\boldmathsymbol{\gamma}^{-1}$ reads
\bea
\label{eq:gamma:bar}
\overline{\boldmathsymbol{\gamma}^{-1}} = \boldmathsymbol{P}^\mathrm{T}\boldmathsymbol{\gamma}^{-1}\boldmathsymbol{P} = 
\begin{pmatrix}
(\boldmathsymbol{\gamma}^{-1})^{\phi\phi}  &  (\boldmathsymbol{\gamma}^{-1})^{\phi\pi}\\
(\boldmathsymbol{\gamma}^{-1})^{\pi\phi} & (\boldmathsymbol{\gamma}^{-1})^{\pi\pi}
\end{pmatrix}\, ,
\eea
the sub-blocks of which are given by
\bea
(\boldmathsymbol{\gamma}^{-1})^{\phi\phi} = \begin{pmatrix}
    (\gamma^{-1})_{11} & (\gamma^{-1})_{13}\\
    (\gamma^{-1})_{31} & (\gamma^{-1})_{33}
\end{pmatrix}, \qquad (\boldmathsymbol{\gamma}^{-1})^{\phi\pi} = \begin{pmatrix}
    (\gamma^{-1})_{12} & (\gamma^{-1})_{14}\\
    (\gamma^{-1})_{32} & (\gamma^{-1})_{34}
\end{pmatrix} \, ,\\
(\boldmathsymbol{\gamma}^{-1})^{\pi\phi} = \begin{pmatrix}
    (\gamma^{-1})_{21} & (\gamma^{-1})_{23}\\
    (\gamma^{-1})_{41} & (\gamma^{-1})_{43}
\end{pmatrix}, \qquad (\boldmathsymbol{\gamma}^{-1})^{\pi\pi} = \begin{pmatrix}
    (\gamma^{-1})_{22} & (\gamma^{-1})_{24}\\
    (\gamma^{-1})_{42} & (\gamma^{-1})_{44}
\end{pmatrix}\, .
\eea 
We now want to diagonalise the quadratic form $\boldmathsymbol{q}^\mathrm{T} \boldmathsymbol{\gamma}^{-1}\boldmathsymbol{q}=\overline{\boldmathsymbol{q}}^\mathrm{T} \overline{\boldmathsymbol{\gamma}^{-1}} \overline{\boldmathsymbol{q}}$, in a way that does not modify its two first entries (such that the argument of the sign functions in \Eq{eq:SxSx:interm0} remains unaffected). This can be done by introducing the new variable $\boldmathsymbol{Q}$ defined as
\bea
\overline{\boldmathsymbol{q}}= 
\begin{pmatrix}
\boldmathsymbol{1} & 0\\
- \left[(\boldmathsymbol{\gamma}^{-1})^{\pi\pi}\right]^{-1} (\boldmathsymbol{\gamma}^{-1})^{\pi\phi}\quad & \boldmathsymbol{1}
\end{pmatrix}
\boldmathsymbol{Q}\, .
\eea
One has
\bea
\boldmathsymbol{q}^\mathrm{T} \boldmathsymbol{\gamma}^{-1}\boldmathsymbol{q} = 
{\boldmathsymbol{Q}}_\phi^{\mathrm{T}}\boldmathsymbol{a} {\boldmathsymbol{Q}}_\phi + 
{\boldmathsymbol{Q}}_\pi^{\mathrm{T}}(\boldmathsymbol{\gamma}^{-1})^{\pi\pi}{\boldmathsymbol{Q}}_\pi \, ,
\eea
where we have introduced
\bea
\label{eq:a:def}
\boldmathsymbol{a} = (\boldmathsymbol{\gamma}^{-1})^{\phi\phi} - (\boldmathsymbol{\gamma}^{-1})^{\phi \pi} [(\boldmathsymbol{\gamma}^{-1})^{\pi\pi}]^{-1} (\boldmathsymbol{\gamma}^{-1})^{\pi \phi}\, 
\eea
and where $\boldmathsymbol{Q}_\phi$ and  $\boldmathsymbol{Q}_\pi$ are the two-dimensional vectors that compose $\boldmathsymbol{Q}$, \ie $\boldmathsymbol{Q}=\begin{pmatrix} \boldmathsymbol{Q}_\phi\\ \boldmathsymbol{Q}_\pi \end{pmatrix}$. Since the Jacobian of the transformation that goes from $\boldmathsymbol{q}$ to $\boldmathsymbol{Q}$ is unity, \Eq{eq:SxSx:interm0} gives rise to 
\bea 
\label{eq:SxSx:interm1}
	\braket{ \hat{\mathcal{S}}_x(\vec{x}_1) \hat{\mathcal{S}}_x(\vec{x}_2)}  = \frac{1}{(2\pi)^2 \sqrt{\det \boldmathsymbol{\gamma}}} &\int \dd Q_1\dd Q_2 \  \textrm{sign}(Q_1) \textrm{sign}(Q_2)\ee^{-\frac{1}{2}\boldmathsymbol{Q}_\phi^{\mathrm{T}} \boldmathsymbol{a}\boldmathsymbol{Q}_\phi}\\
	& \times \int \dd Q_3\dd Q_4\ee^{-\frac{1}{2}\boldmathsymbol{Q}_\pi^{\mathrm{T}} (\boldmathsymbol{\gamma}^{-1})^{\pi\pi}\boldmathsymbol{Q}_\pi}\, .
\eea
The integral over $\boldmathsymbol{Q}_\pi$ is a simple Gaussian integral and can be readily performed. Upon splitting the integral over $Q_1$ and $Q_2$ according to their sign, one then finds
\bea 
\label{eq:SxSx:interm2}
	\braket{ \hat{\mathcal{S}}_x(\vec{x}_1) \hat{\mathcal{S}}_x(\vec{x}_2)}  =& \frac{1}{2\pi \sqrt{\det \boldmathsymbol{\gamma}\det (\boldmathsymbol{\gamma}^{-1})^{\pi\pi}  }} \left[\int \dd Q_1\dd Q_2 \ee^{-\frac{1}{2}\boldmathsymbol{Q}_\phi^{\mathrm{T}} \boldmathsymbol{a}\boldmathsymbol{Q}_\phi}
	\right. \\ & \left.
	-2 \int_{0}^\infty \dd Q_1 \int_{-\infty}^0 \dd Q_2 \ee^{-\frac{1}{2}\boldmathsymbol{Q}_\phi^{\mathrm{T}} \boldmathsymbol{a}\boldmathsymbol{Q}_\phi}
	-2 \int^{0}_{-\infty} \dd Q_1 \int^{\infty}_0 \dd Q_2 \ee^{-\frac{1}{2}\boldmathsymbol{Q}_\phi^{\mathrm{T}} \boldmathsymbol{a}\boldmathsymbol{Q}_\phi}\right]
	\, .
\eea
The first integral is again a simple Gaussian integral, while the second and third integrals are equal because of the invariance of the problem under exchanging $\vec{x}_1$ and $\vec{x}_2$. It can be computed by  first integrating over $Q_2$ and then over $Q_1$:
\bea
\int_{0}^\infty \dd Q_1 \int_{-\infty}^0 \dd Q_2 \ee^{-\frac{1}{2}\boldmathsymbol{Q}_\phi^{\mathrm{T}} \boldmathsymbol{a}\boldmathsymbol{Q}_\phi}&=
\int_{0}^\infty \dd Q_1 \int_{-\infty}^0 \dd Q_2 \ee^{-\frac{1}{2}(a_{11} Q_1^2 + a_{22} Q_2^2 + 2 a_{12} Q_1 Q_2)}\\
& = \int_{0}^\infty \dd Q_1 \ee^{\left(-\frac{a_{11}}{2}+\frac{ a_{12}^2}{2 a_{22}}\right)Q_1^2} \int_{-\infty}^0 \dd Q_2  \ee^{-\frac{a_{22}}{2}\left(Q_2+\frac{a_{12}}{a_{22}}Q_1\right)^2}\\
& = \int_{0}^\infty \dd Q_1 \ee^{\left(-\frac{a_{11}}{2}+\frac{ a_{12}^2}{2 a_{22}}\right)Q_1^2} \sqrt{\frac{\pi}{2 a_{22}}} \left[1+\mathrm{erf}\left(\frac{a_{12}}{\sqrt{2 a_{22}}} Q_1\right)\right]\\
&= \frac{\frac{\pi}{2}+\mathrm{arctan}\left(\frac{a_{12}}{a_{11}a_{22}-a_{12}^2}\right)}{\sqrt{a_{11}a_{22}-a_{12}^2}}\, .
\eea
Combining the above results, one obtains
\bea 
\label{eq:SxSx:interm3}
	\braket{ \hat{\mathcal{S}}_x(\vec{x}_1) \hat{\mathcal{S}}_x(\vec{x}_2)}  =& -\frac{2}{\pi}\sqrt{\frac{\det \boldmathsymbol{\gamma}^{-1}}{\det(\boldmathsymbol{\gamma}^{-1})^{\pi\pi} \det \boldmathsymbol{a}}}\mathrm{arctan}\left(\frac{a_{12}}{a_{11}a_{22}-a_{12}^2}\right)	\, .
\eea
This expression can be further simplified as follows. Using the formula for determinants of block matrices in \Eq{eq:gamma:bar}, one has
\bea
\det \boldmathsymbol{\gamma}^{-1} = \det  \overline{\boldmathsymbol{\gamma}^{-1}} & =  \det (\boldmathsymbol{\gamma}^{-1})^{\pi\pi} \det \left\{(\boldmathsymbol{\gamma}^{-1})^{\phi\phi} - (\boldmathsymbol{\gamma}^{-1})^{\phi \pi} [(\boldmathsymbol{\gamma}^{-1})^{\pi\pi}]^{-1} (\boldmathsymbol{\gamma}^{-1})^{\pi \phi} \right\}\\
& =  \det (\boldmathsymbol{\gamma}^{-1})^{\pi\pi} \det \boldmathsymbol{a}\, ,
\eea
where we have recognised the matrix $\boldmathsymbol{a}$ defined in \Eq{eq:a:def}. Upon replacing $\det\boldmathsymbol{a}= a_{11} a_{22}-a_{12}^2$ in \Eq{eq:SxSx:interm3} by $\det \boldmathsymbol{\gamma}^{-1}/\det (\boldmathsymbol{\gamma}^{-1})^{\pi\pi}$, one finally obtains
\bea 
	\braket{\hat{\mathcal{S}}_x(\vec{x}_1) \hat{\mathcal{S}}_x(\vec{x}_2)} = 
	- \frac{2}{\pi} \arctan\left[ a_{12} (a_{11} a_{22} - a_{12}^2)^{-1/2}\right]\, ,
\eea 
which is the equation used in the main text.

The other spin correlators are more straightforward to evaluate. One has
\bea
	\braket{\hat{\mathcal{S}}_x(\vec{x}_1) \hat{\mathcal{S}}_z(\vec{x}_2)}& = \frac{1}{(2\pi)^2 \sqrt{\det \boldmathsymbol{\gamma}}} \int \dd\boldmathsymbol{q} \left[-\pi \delta(q_1)\right] \delta(q_2) \textrm{sign}(q_3))\ee^{-\frac{1}{2}\boldmathsymbol{q}^{\mathrm{T}} \boldmathsymbol{\gamma}^{-1}\boldmathsymbol{q}}\\
	& = \frac{-1}{4\pi \sqrt{\det \boldmathsymbol{\gamma}}} \int \dd q_3 \dd q_4 \textrm{sign}(q_3) e^{-\frac{1}{2}q_3 (\gamma^{-1})_{33} q_3 -\frac{1}{2} q_4 (\gamma^{-1})_{44} q_4 - q_4 (\gamma^{-1})_{43} q_3}\\
	& = \frac{-1}{4\pi \sqrt{\det \boldmathsymbol{\gamma}}} \sqrt{\frac{2\pi}{(\gamma^{-1})_{44}}}\int \dd q_3 \textrm{sign}(q_3) e^{-\frac{1}{2}q_3 (\gamma^{-1})_{33} q_3 + \frac{\left[(\gamma^{-1})_{43} q_3\right]^2}{2(\gamma^{-1})_{44}} }\\
	& = 0
\eea
since the last integrand is an odd function, to be integrated over the real line. One finally has
\bea
	\braket{\hat{\mathcal{S}}_z(\vec{x}_1) \hat{\mathcal{S}}_z(\vec{x}_2)} & = \frac{1}{(2\pi)^2 \sqrt{\det \boldmathsymbol{\gamma}}} \int \dd \boldmathsymbol{q} \pi^2 \delta(q_1) \delta(q_2)\delta(q_3) \delta(q_4)e^{-\frac{1}{2} \boldmathsymbol{q}^{\mathrm{T}} \boldmathsymbol{\gamma}^{-1} \boldmathsymbol{q}}\\
	& = \frac{1}{4 \sqrt{\det \boldmathsymbol{\gamma}}}\,.
\eea
\section{Larsson spin correlators}
\label{sec:Larsson:Spin:Correlators}
\subsection*{Intermediate analytical expressions}
By plugging \Eqs{eq:Wigner:Gauss} and~\eqref{eq:Weyl:Larsson} into \Eq{eq:mean:weyl}, one obtains explicit expressions for the two-point correlation functions of the Larsson spin operators, involving a double sum of double integrals. One of these integrals [say the one over $\tilde{\phi}_R(\vec{x}_2)$] can be performed in terms of the error function, and one obtains
\bea
	\braket{\hat{S}^\ell_z(\vec{x}_1) \hat{S}^\ell_z(\vec{x}_2)} & =  \frac{1}{2\pi \sqrt{\det \boldmathsymbol{\gamma} \det(\boldmathsymbol{\gamma}^{-1})^{\pi\pi}}}\sum_{n,m = -\infty}^{\infty} (-1)^{n+m} \mathcal{Z}_{n,m}(\vec{x}_1,\vec{x}_2),\\
	\braket{\hat{S}^\ell_x(\vec{x}_1) \hat{S}^\ell_x(\vec{x}_2)} & = \frac{1}{2\pi \sqrt{\det \boldmathsymbol{\gamma}\det(\boldmathsymbol{\gamma}^{-1})^{\pi\pi}}}  \sum_{n,m = -\infty}^\infty  \mathcal{X}_{n,m}(\vec{x}_1,\vec{x}_2),\\
	\braket{\hat{S}^\ell_x(\vec{x}_1) \hat{S}^\ell_z(\vec{x}_2)} & = 0\,.
\eea
The functions $\mathcal{Z}_{n,m}$ and $\mathcal{X}_{n,m}$ are defined as
\bea
\label{eq:Z:X:def}
	\mathcal{Z}_{n,m}(\vec{x}_1,\vec{x}_2) & = \sqrt{\frac{\pi}{2 a_{22}}}\int_{n\ell}^{(n+1)\ell} \dd\phi e^{-\frac{1}{2}\phi^2 \left(a_{11} - \frac{a_{12}^2}{a_{22}} \right)}\\
	& \quad  \times \left\lbrace \textrm{erf}\left[ \frac{a_{12} \phi + a_{22}(m+1)\ell}{\sqrt{2a_{22}}} \right] - \textrm{erf}\left[ \frac{a_{12} \phi + a_{22} m \ell}{\sqrt{2a_{22}}} \right]\right\rbrace\\
	\mathcal{X}_{n,m} (\vec{x}_1,\vec{x}_2) & = \int_{\frac{\ell}{2} + 2nl}^{\frac{\ell}{2}+(2n+1)\ell} \dd\phi \sum_{\epsilon_1,\epsilon_2=-1,1} \sqrt{\frac{\pi}{2 a_{22}}}  e^{-\frac{\phi^2}{2} \left(a_{11} - \frac{a_{12}^2}{a_{22}} \right) + \frac{i\ell \phi}{a_{22}}\left(a_{12}\tilde{a}_{y} - a_{22} \tilde{a}_1 \right)- \frac{\ell^2\tilde{a}^2_y}{2a_{22}}- \frac{\ell^2}{2} \boldmathsymbol{\epsilon}^{\mathrm{T}}\cdot (\boldmathsymbol{\gamma}^{-1})^{\pi\pi}\cdot\boldmathsymbol{\epsilon}}\\
	& \quad  \times \left\lbrace \textrm{erf}\left[ \frac{a_{12} \phi +i\ell\tilde{a}_2 + a_{22}\left(2m+\frac{3}{2}\right)\ell}{\sqrt{2a_{22}}} \right] - \textrm{erf}\left[ \frac{a_{12} \phi + i\ell\tilde{a}_2 + a_{22} \left(2m+\frac{1}{2}\right) \ell}{\sqrt{2a_{22}}} \right]\right\rbrace,
\eea
where we recall that the matrix $\boldmathsymbol{a}$ was introduced in \Eq{eq:a:def}, and where the integration variable $\phi$ physically correspond to $\tilde{\phi}_R(\vec{x})$. We have also introduced the vector $\boldmathsymbol{\epsilon}=(\epsilon_1,\epsilon_2)^{\mathrm{T}}$, and the quantities $\tilde{a}_1$ and $\tilde{a}_2$ defined as
\bea
    \tilde{a}_1  = & \frac{1}{2} \bigg\lbrace (\gamma^{-1})_{12}[(\gamma^{-1})^{\pi\pi}]^{-1}_{11} \epsilon_x + (\gamma^{-1})_{12}[(\gamma^{-1})^{\pi\pi}]^{-1}_{12} \epsilon_y  \\ & + (\gamma^{-1})_{14}[(\gamma^{-1})^{\pi\pi}]^{-1}_{21} \epsilon_x + (\gamma^{-1})_{14}[(\gamma^{-1})^{\pi\pi}]^{-1}_{22} \epsilon_y \bigg\rbrace ,\\
    \tilde{a}_2  = & \frac{1}{2} \bigg\lbrace (\gamma^{-1})_{32}[(\gamma^{-1})^{\pi\pi}]^{-1}_{11} \epsilon_x + (\gamma^{-1})_{32}[(\gamma^{-1})^{\pi\pi}]^{-1}_{12} \epsilon_y + \\ & + (\gamma^{-1})_{34}[(\gamma^{-1})^{\pi\pi}]^{-1}_{21} \epsilon_x + (\gamma^{-1})_{34}[(\gamma^{-1})^{\pi\pi}]^{-1}_{22} \epsilon_y \bigg\rbrace .
\eea
In general, the remaining integral and double sum need to be carried out numerically. However, more analytical insight can be gained in the small-$\ell$ and the large-$\ell$ limits.
\subsection*{Small-$\ell$ limit}
When $\ell\ll 1$, the functions $\mathcal{Z}_{n,m}$ and $\mathcal{X}_{n,m}$ involve differences of error functions evaluated at nearby points. They can therefore be Taylor expanded as follows
\bea
	\textrm{erf}\left[b_1 + b_2 (m+1)\ell\right] - \textrm{erf}(b_1 + b_2 m\ell) \simeq b_2 \ell \frac{\dd}{\dd(b_2 m\ell)} \textrm{erf}(b_1 + b_2 m\ell) = \frac{2b_2 \ell}{\sqrt{\pi}} e^{-(b_1 + b_2 m \ell)^2}\, ,
\eea
and
\bea
\text{erf}\left[b_1 + b_2\left(2m+\frac{3}{2}\right)\ell\right] - \text{erf}\left[b_1 + b_2\left(2m+\frac{1}{2}\right)\ell\right]& \simeq \ell b_2 \frac{\dd}{\dd (2mb_2\ell)} \text{erf} \left[b_1 + b_2 \ell (2m+1/2)\right]\\ & = \frac{2 b_2 \ell}{\sqrt{\pi}} \ee^{-\left[b_1 + b_2 \ell \left(2m+\frac{1}{2}\right)\right]^2}\, .
\eea
The remaining integrals in $\mathcal{Z}_{n,m}$ and $\mathcal{X}_{n,m}$ become Gaussian integrals and can thus be performed analytically. Similarly, the sums over $n$ and $m$ can be approximated by Riemann integrals upon introducing $x = n\ell$ and $y = m\ell$ and by noticing that
\bea
    \sum_{n,m = -\infty}^\infty \ell^2 g(n\ell,m\ell) \underset{\ell\ll 1}{\simeq} \int_{-\infty}^\infty \dd x \dd y g\left(x,y\right)\,,
\eea
if $g$ is a sufficiently smooth function.

In the particular case of interest to us, since $g(n,m)$ is Gaussian, one can compute its Riemann integral right away. This finally gives rise to
\bea
\braket{\hat{S}^\ell_z(\vec{x}_1) \hat{S}^\ell_z(\vec{x}_2)} \simeq e^{-\frac{\pi^2 (a_{11} + a_{22} - 2 a_{12})}{2\ell^2 (a_{11} a_{22} - a_{12}^2)}}\underset{\ell\to 0}{\longrightarrow} 0
\eea
and
\bea
\label{eq:app:SxSx:small:ell}
\braket{\hat{S}^\ell_x(\vec{x}_1) \hat{S}^\ell_x(\vec{x}_2)} \simeq \frac{1}{4} \sum_{\boldmathsymbol{\epsilon}} e^{ \frac{\ell^2}{2}
\left\lbrace \frac{a_{11} \tilde{a}_2^2 + a_{22} \tilde{a}_1^2 - 2 a_{12} \tilde{a}_1 \tilde{a}_2 }{a_{11} a_{22} - a_{12}^2} +  \boldmathsymbol{\epsilon}^T\cdot [(\gamma^{-1})^{\pi\pi}]\cdot\boldmathsymbol{\epsilon} \right\rbrace } \underset{\ell\to 0}{\longrightarrow} 1\, .
\eea 
\subsection*{Large-$\ell$ limit}
When $\ell\to\infty$, in the expression for $\hat{S}_x^\ell (\vec{x})$ given in \Eq{eq:Sz:Larsson}, one can see that only the term with $n=0$ has a non-empty integration domain. However, this term involves the field eigenstate $\vert\infty\rangle$, which necessarily vanishes when evaluated on a normalised state. This implies that 
\bea
\braket{\hat{S}^\ell_x(\vec{x}_1) \hat{S}^\ell_x(\vec{x}_2)}
\underset{\ell\to \infty}{\longrightarrow} 0\, ,
\eea 
which can be further checked by noticing that for all $m$, the two error functions appearing in the expression of $\mathcal{X}_{n,m}$ given in \Eq{eq:Z:X:def} are evaluated at the same point (\ie either $+\infty$ or $-\infty$), hence the difference always vanishes.

For $\hat{S}_z^\ell (\vec{x})$, when evaluating the expression given in \Eq{eq:Sz:Larsson} in the limit $\ell\to\infty$, only the terms $n=-1$ and $n=0$ remain, which leads to
\bea
\hat{S}_z^\ell (\vec{x}) & \underset{\ell\to \infty}{\longrightarrow}
\int_{-\infty}^{0} \dd\tilde{\phi}_R(\vec{x}) \ket{\tilde{\phi}_R(\vec{x})}{\bra{\tilde{\phi}_R(\vec{x})}}
+\int^{\infty}_{0} \dd\tilde{\phi}_R(\vec{x}) \ket{\tilde{\phi}_R(\vec{x})}{\bra{\tilde{\phi}_R(\vec{x})}}\\ &
=\int_{-\infty}^{\infty} \dd\tilde{\phi}_R(\vec{x}) \ket{\tilde{\phi}_R(\vec{x})}{\bra{\tilde{\phi}_R(\vec{x})}}\, .
\eea 
This formula coincides with the one for the $\hat{\mathcal{S}}_x$ component of the GKMR operator. Indeed, by plugging \Eq{eq:E:O:def} into \Eq{eq:GKMR:def}, one obtains
\bea
\hat{\mathcal{S}}_x (\vec{x}) & = \frac{1}{2}\int_{0}^{\infty} \dd\tilde{\phi}_R(\vec{x}) \left[\ket{\tilde{\phi}_R(\vec{x})}{\bra{\tilde{\phi}_R(\vec{x})}}-\ket{-\tilde{\phi}_R(\vec{x})}{\bra{-\tilde{\phi}_R(\vec{x})}}\right]
\\ & =\int_{-\infty}^{\infty} \dd\tilde{\phi}_R(\vec{x}) \ket{\tilde{\phi}_R(\vec{x})}{\bra{\tilde{\phi}_R(\vec{x})}}\, ,
\eea 
where the change of integration variable $\tilde{\phi}_R(\vec{x})\to-\tilde{\phi}_R(\vec{x})$ has been performed in the second term. One thus has 
\bea
\hat{S}_z^\ell (\vec{x}) & \underset{\ell\to \infty}{\longrightarrow}\hat{\mathcal{S}}_x(\vec{x}) ,
\eea 
hence the two-point function of the $\hat{\mathcal{S}}_x^\ell (\vec{x})$ operator is given by \Eq{eq:SxSx} in the limit $\ell\to\infty$.
\section{Additional figures}
\label{app:Additional:Figures}
In this appendix, we provide additional figures, which are not directly relevant to the discussion presented in the main text, but which nonetheless complete our parameter-space exploration. 

\begin{figure}
    \centering
    \includegraphics[width=0.99\textwidth]{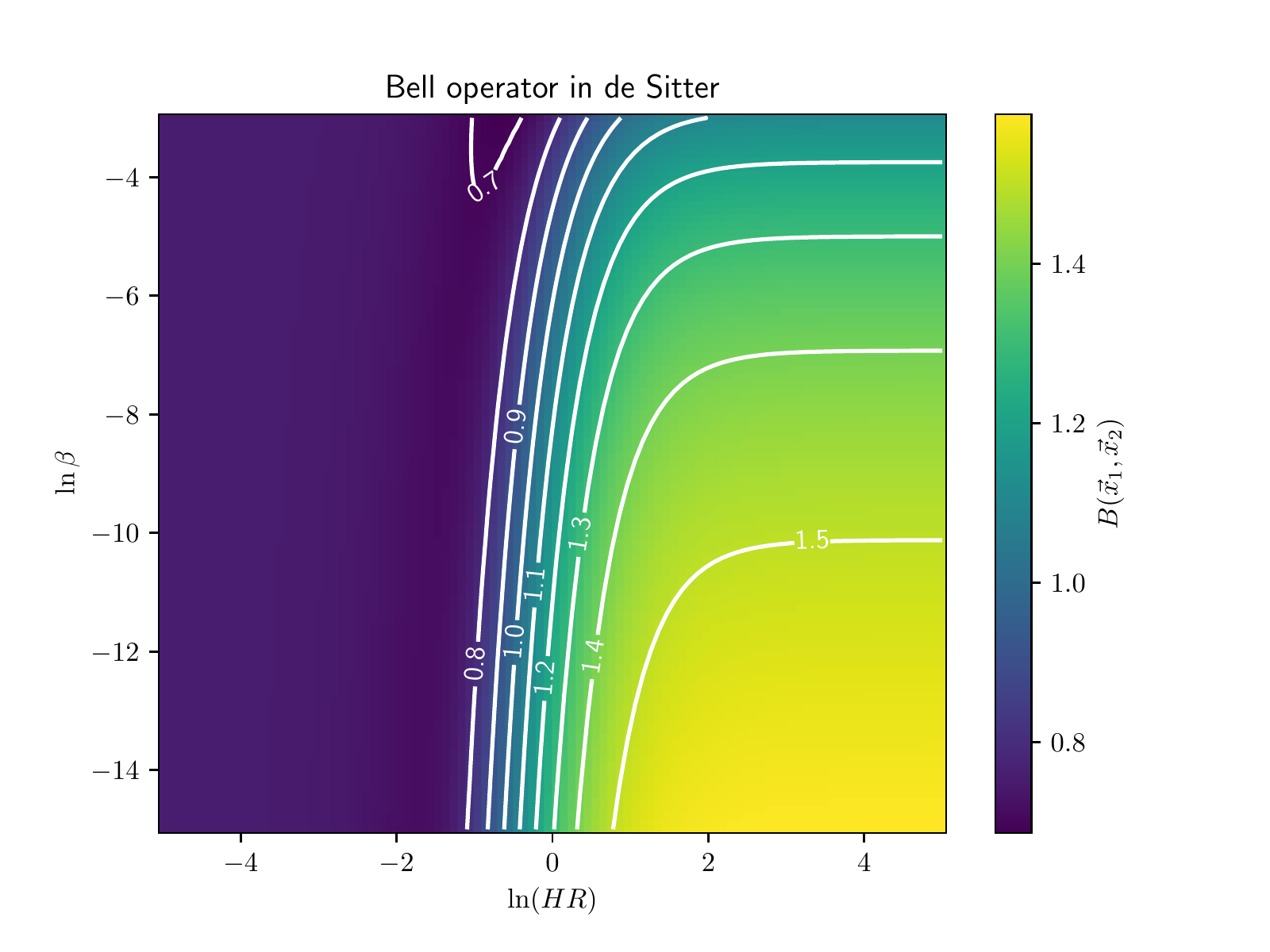}
    \caption{Expectation value of the GKMR Bell operator in the Bunch-Davies vacuum of the de-Sitter space-time, as a function of the parameters $\beta$ and $HR$. The colour encodes the value of $B$, and a few contour lines are displayed in white. The UV regulator is set to $\delta = 0.01$ and $\alpha = d/R$ is set to the minimum $\alpha = 2(\delta + 1)$. 
    }
    \label{fig:dS_beta_HR}
\end{figure}

\begin{figure}
    \centering
    \includegraphics[width=0.99\textwidth]{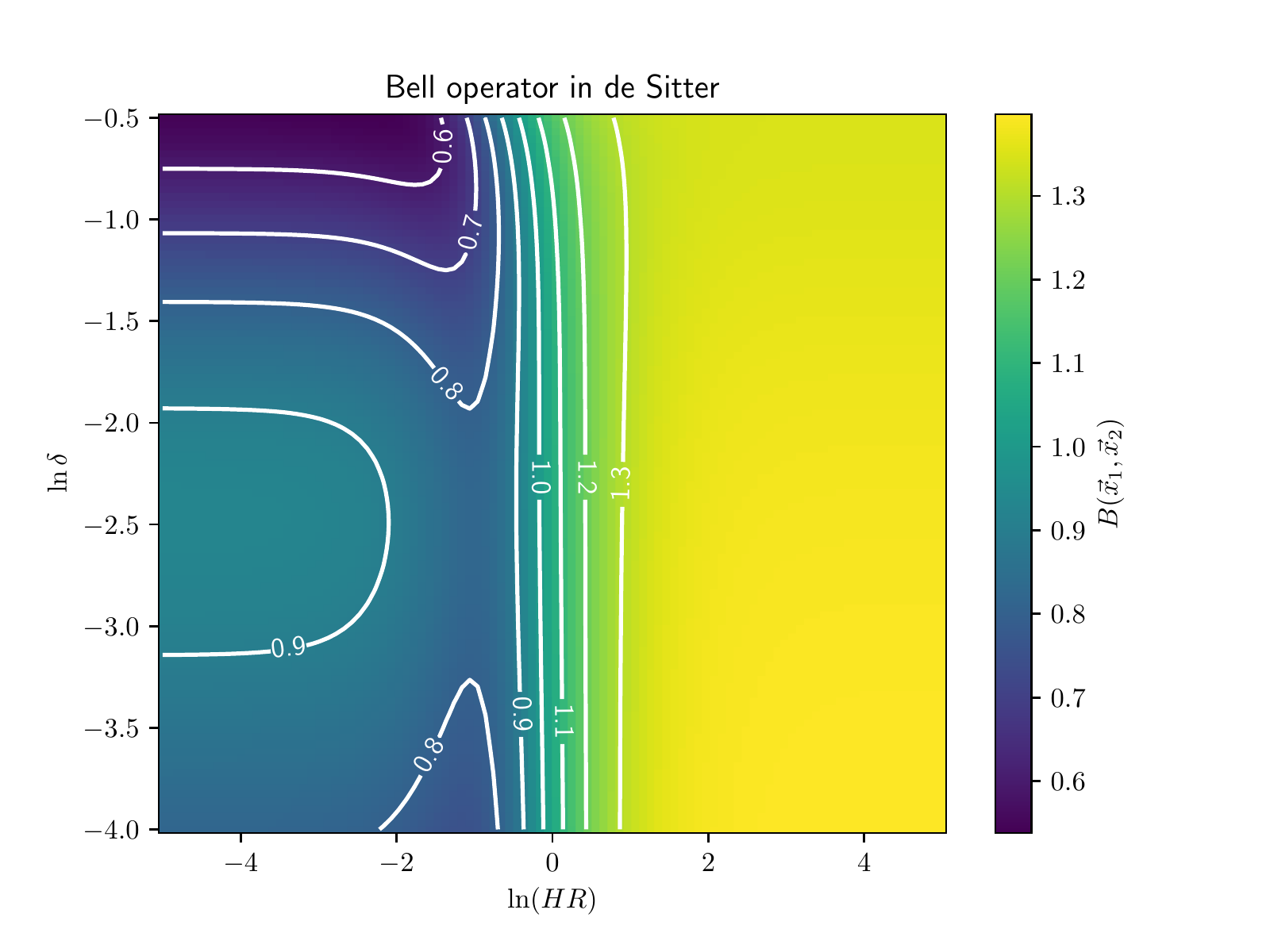}
    \caption{Expectation value of the GKMR Bell operator in the Bunch-Davies vacuum of the de-Sitter space-time, as a function of the parameters $\beta$ and $HR$. The colour encodes the value of $B$, and a few contour lines are displayed in white. The IR regulator is set to $\beta = 10^{-3}$ and $\alpha = d/R$ is set to the minimum $\alpha = 2(\delta + 1)$.
    }
    \label{fig:dS_delta_HR}
\end{figure}

\begin{figure}
    \centering
    \includegraphics[width=0.99\textwidth]{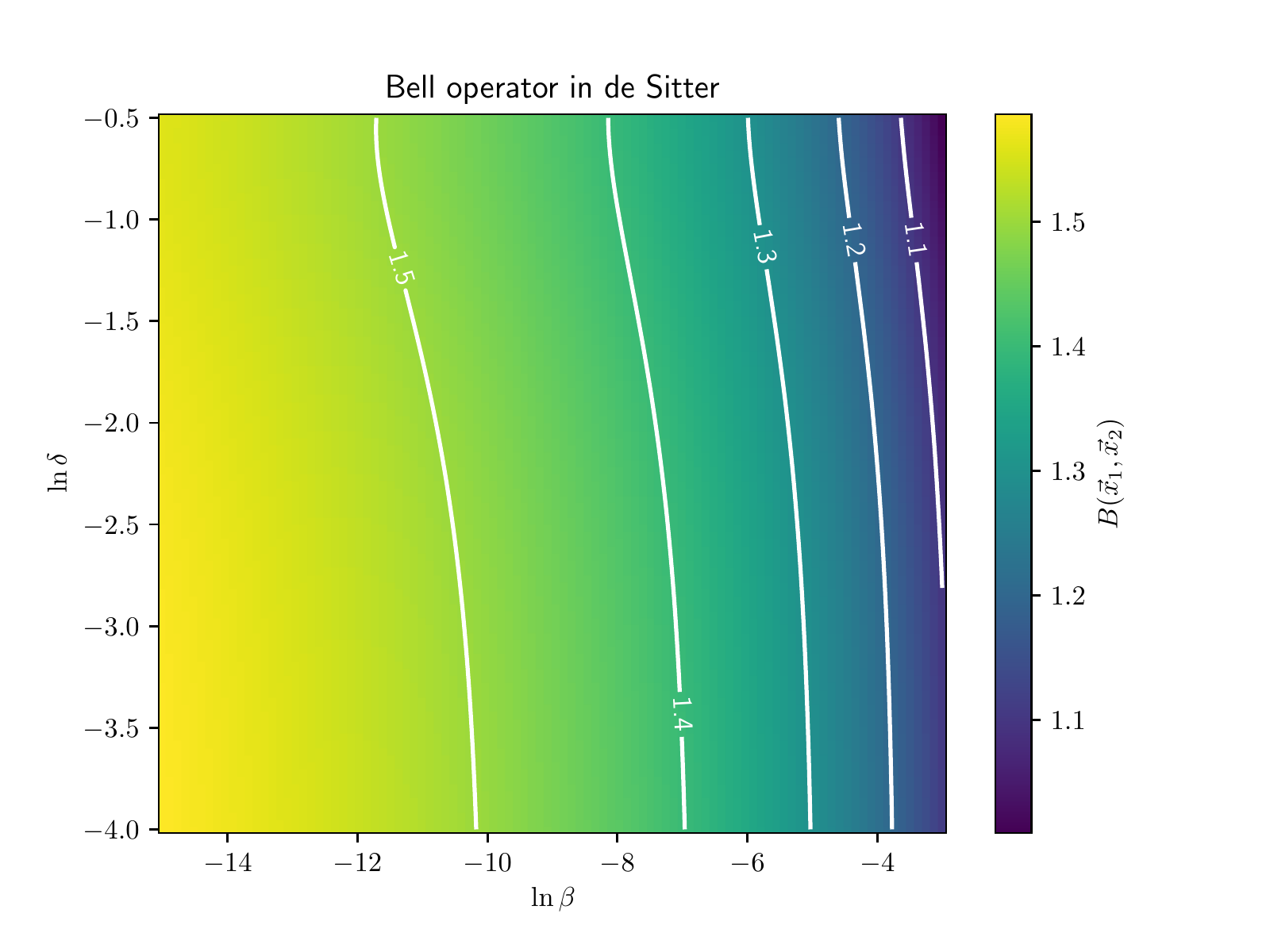}
    \caption{Expectation value of the GKMR Bell operator in the Bunch-Davies vacuum of the de-Sitter space-time, as a function of the parameters $\beta$ and $HR$. The colour encodes the value of $B$, and a few contour lines are displayed in white. The size of the patch is set to $HR = 10^3$ and $\alpha = d/R$ is set to the minimum $\alpha = 2(\delta + 1)$.}
    \label{fig:dS_beta_delta}
\end{figure}

\bibliographystyle{JHEP}
\bibliography{deSitterBell}

\end{document}